\documentstyle[graphicx,psfig,epsfig,amsmath,amssymb,enumerate]{mn}

\title[Bayesian galaxy shape measurement]
{Bayesian Galaxy Shape Measurement for Weak Lensing Surveys -II. Application to Simulations}
\author[T. Kitching et al. ]{T.D. Kitching\thanks{tdk@astro.ox.ac.uk}$^{1}$, L. Miller$^{1}$, 
C.E. Heymans$^{2}$, L. van Waerbeke$^{2}$, A.F. Heavens$^{3}$
\\  
$^{1}$Department of Physics, Oxford University, Keble Road, Oxford, OX1 3RH, U.K.\\
$^{2}$University of British Columbia, Department of Physics and Astronomy, 6224 Agricultural Road, Vancouver, B.C. V6T 1Z1,Canada\\
$^{3}$SUPA\thanks{Scottish Universities Physics Alliance}, Institute for Astronomy, University of Edinburgh, Royal Observatory, Blackford Hill, Edinburgh, EH9 3HJ, UK}

\newcommand{\be}{\begin{equation}}
\newcommand{\ee}{\end{equation}}
\newcommand{\ba}{\begin{eqnarray}}
\newcommand{\ea}{\end{eqnarray}}
\newcommand{\nn}{\nonumber \\}
\def\gs{\mathrel{\raise1.16pt\hbox{$>$}\kern-7.0pt %
\lower3.06pt\hbox{{$\scriptstyle \sim$}}}}         %
\def\ls{\mathrel{\raise1.16pt\hbox{$<$}\kern-7.0pt %
\lower3.06pt\hbox{{$\scriptstyle \sim$}}}}         %

\begin{document}


\maketitle


\begin{abstract}
In this paper we extend the Bayesian model fitting shape measurement method presented in 
Miller et al. (2007) and use the method to estimate the shear from the Shear TEsting Programme 
simulations (STEP). The method uses a fast model fitting 
algorithm which uses realistic galaxy profiles and analytically marginalises over the position and 
amplitude of the model by doing the model fitting in Fourier space. This is used to 
find the full posterior probability in ellipticity. The shear is then estimated in a Bayesian 
way from this posterior probability surface. The Bayesian estimation allows 
measurement bias arising from the presence of random noise to be removed. 
In this paper we introduce an iterative algorithm that can be used to estimate 
the intrinsic ellipticity prior and show that this is accurate and stable.

We present results using the STEP parameterisation which relates the input shear $\gamma^T$ to the 
estimated shear $\gamma^M$ by introducing a bias $m$ and an offset $c$: 
$\gamma^M-\gamma^T= m\gamma^T + c$.
By using the method to estimate the shear from the STEP1 simulations we find the method to have 
a shear bias of $m\sim 5\times 10^{-3}$ and a variation in shear offset with PSF type of 
$\sigma_c\sim 2\times 10^{-4}$. These values are smaller than for any method presented in 
the STEP1 publication that behaves linearly with shear. 
Using the method to estimate the shear from the STEP2 simulations we find than the shear 
bias and offset are $m\sim 2\times 10^{-3}$ and $c\sim -7\times 10^{-4}$ 
respectively. 
In addition we find that the 
bias and offset are stable to changes in magnitude and size of the galaxies. 
Such biases should yield any cosmological constraints from future 
weak lensing surveys robust to systematic effects in shape measurement.

Finally we present an alternative to the STEP parameterisation by using a    
Quality factor that relates the intrinsic shear variance in a simulation 
to the variance in shear that is measured and show that the method presented has an 
average of $Q\gs 100$ which is at least a factor of $10$ times better than other shape measurement methods.
\end{abstract}

\begin{keywords}
Gravitational lensing - Methods: numerical, statistical, data analysis - Cosmology : observation
\end{keywords}

\section{Introduction}
It has been shown that weak lensing 
has the potential to become one of our most powerful cosmological probes (see 
Munshi et al., 2007 for a recent review of weak lensing; DETF, 
Albrecht et al., 2007; Peacock et al., 2007). By using redshift 
and weak lensing information 3D weak lensing techniques have been developed that are 
particularly sensitive to the dark energy equation of state (for example Heavens et al., 2006; 
Taylor et al., 2007). Since the promise of weak lensing 
is now firmly established one must begin to focus on refining the technique and addressing 
systematic issues. 

The determination of galaxy shape, and the inference of shear across an ensemble of galaxies 
for use in weak lensing is a challenging problem with a rich history. 
However 
recent studies of weak lensing systematic effects (e.g. Kitching et al., 2008a; Amara \& 
Refregier, 2007) have shown that in order to fully utilise future weak lensing 
surveys (e.g. DUNE, Refregier et al., 2006; Pan-STARRS, Kaiser et al, 2002; 
SNAP, Kim et al., 2002; LSST, Tyson et al., 2003) 
in the determination of cosmological parameters, such as  
the equation of state of dark energy, the bias in the estimated shear as a result 
of any difference between a galaxy's true shape and the measured shape needs to be 
$\Delta e/e < 10^{-3}$.
Currently-used methods, tested on simulations, have at best a $10^{-2}$ bias (Heymans et al., 2006). 

In this paper we expand upon and apply to simulations the new shape measurement method 
{\sc lensfit}\footnote{For further information, and to download publically available code,
 please go to http://www.physics.ox.ac.uk/lensfit.html} presented in 
Miller et al. (2007), a method which uses realistic galaxy profiles 
and fits these models to images using a fast fitting algorithm. The fast model fitting 
approach allows the entire posterior probability surface in ellipticity to be calculated. By 
including a prior the estimation of the shear can then be done in a fully Bayesian way. It was 
shown that a Bayesian estimator should be unbiased, given that realistic models and an 
accurate and correct intrinsic ellipticity prior are used. A new bias 
was discovered  as a result of assuming that the prior is centred on zero-shear, which must be 
assumed given no knowledge of the intrinsic ellipticity distribution, but it was shown that this 
bias can be exactly corrected to first order within the Bayesian formalism. 

The simulations analysed in this paper are the publically available simulations from 
STEP (Shear TEsting Programme)\footnote{http://www.physics.ubc.ca/$\sim$heymans/step.html}. The 
currently published STEP papers present the accuracy with which currently available shape 
measurement methods can recover the input shear from simulations of varying complexity. 
STEP1 (Heymans et al., 2006) used simulated galaxies 
which consist of a de Vaucouleurs bulge plus an exponential disk (in varying degrees), these are 
provided in sets of images with varying PSFs and shear, there are $64$ images covering 
$5$ shear values for each of $5$ PSF types.  
In STEP2 Massey et al. (2007) used shapelet generated galaxies and exponential galaxies, these are 
provided in sets of $128$ images for each of $6$ different PSFs. 

In Miller et al. (2007) the results shown were for individual galaxy ellipticities, in this paper we
present results for shear. The results presented compare the estimated shear found using 
our technique with the known input shear of the simulations. Sections 
\ref{Overview of lensfit} and \ref{Estimation of the Prior} review the shape measurement method, 
as well as extend the development by introducing a new way to determine the prior 
intrinsic ellipticity distribution from data. In Section \ref{Results of tests on simulations} 
we describe the simulations in more detail and present the results from the STEP1 and STEP2 
respectively. We present a new way to characterise a shape measurement methods 
performance in Section \ref{Beyond m and c values}. 
Discussion and conclusions will be presented in Section \ref{Conclusion}.

\section{Overview of lensfit}
\label{Overview of lensfit}
This Section presents an overview of the {\sc lensfit} shape measurement method, for an full 
description see Miller et al. (2007). 

The method presented here combines two innovations in the shape measurement problem. Firstly the 
shear estimation is done in a fully \emph{Bayesian} way, given a likelihood in ellipticity 
generated by some procedure and a prior on ellipticity it is possible to construct a shear 
estimator that is in principle unbiased. 
Secondly we use \emph{realistic} galaxy profiles to generate a full 
posterior probability surface in ellipticity.  

Note that the Bayesian shear estimation formalism can be applied to \emph{any} shear 
measurement method
that can produce a full likelihood surface in ellipticity. Similarly the fast model fitting algorithm 
could be applied to any choice of model. 

\subsection{Overview of Bayesian Galaxy Shape Measurement}
\label{Overview of Bayesian Galaxy Shape Measurement}

For each galaxy a (Bayesian) posterior probability in ellipticity can be generated
\be
p_i(\bmath{e} | \bmath{y}_i) = 
\frac{
{\mathcal P}\left (\bmath{e}\right ) {\mathcal L}\left (\bmath{y}_i | \bmath{e}\right )}
{\int{\mathcal P}\left (\bmath{e}\right ) {\mathcal L}\left (\bmath{y}_i | \bmath{e}\right ) d\bmath{e}}
\ee
where ${\mathcal P}\left (\bmath{e}\right )$ is the ellipticity prior
probability distribution and
${\mathcal L}\left (\bmath{y}_i | \bmath{e}\right )$
is the likelihood of obtaining the $i^{\rm th}$ set of data values $\bmath{y}_i$
given an intrinsic ellipticity $\bmath{e}$.

We would hope that by considering the summation over the data 
the true distribution of intrinsic ellipticities can be obtained from the data 
\be
\langle
\frac{1}{N}
\sum_i p_i(\bmath{e} | \bmath{y}_i) 
\rangle
=
\int d\bmath{y} \frac
{{\mathcal P}\left (\bmath{e}\right ) {\mathcal L}\left (\bmath{y} | \bmath{e}\right )}
{\int{\mathcal P}\left (\bmath{e}\right ) {\mathcal L}\left (\bmath{y} | \bmath{e}\right ) d\bmath{e}}
\int f(\bmath{e}) \epsilon(\bmath{y}|\bmath{e}) d\bmath{e}
\ee
where $\epsilon(\bmath{y}|\bmath{e})$ is the probability distribution 
for the data $\bmath{y}$ given an ellipticity $\bmath{e}$ and $f(\bmath{e})$ is the true (intrinsic) 
ellipticity distribution.
On the right-hand-side (RHS) we are integrating over the probability distributions to
obtain the expectation value of the
summed posterior probability distribution for the sample.
This will be achieved under the conditions that 
$\epsilon(\bmath{y} | \bmath{e}) = {\mathcal L}\left (\bmath{y} | \bmath{e}\right )$
and 
${\mathcal P}\left (\bmath{e}\right ) = f\left (\bmath{e}\right )$ (assuming the
likelihood is normalised,
$\int {\mathcal L}\left (\bmath{y} | \bmath{e}\right) d\bmath{y} = 1$)
from which we obtain
\be
\label{cent}
\langle
\frac{1}{N}
\sum_i p_i(\bmath{e} | \bmath{y}) 
\rangle
= {\mathcal P}\left (\bmath{e} \right )
= f(\bmath{e}).
\ee
This is the equation that highlights the essence of the 
Bayesian shape measurement method, given a prior that matches
the intrinsic distribution of ellipticities the estimated posterior probability should be 
unbiased. 

It 
may appear at first that having an accurate and correct measure of the prior distribution before 
estimating the ellipticity of galaxies may represent \emph{petitio principii} 
however this was partially addressed 
in Miller et al. (2007) and we extend and validate the issue of creating the prior in 
Section \ref{Estimation of the Prior}.  

Throughout this paper we assume a galaxy's ellipticity $\bmath{e}$ is defined by relating 
the axial ratio $\beta$ 
and orientation $\phi$ of the galaxy via  
\be 
\label{axialr}
\left( \begin{array}{c}
e_1\\
e_2 \end{array}\right)=\frac{1-\beta}{1+\beta}\left( \begin{array}{c}
\cos[2\phi]\\
\sin[2\phi]\end{array}\right).
\ee
The ellipticity can be 
related to the intrinsic galaxy ellipticity $\bmath{e^s}$ in the weak lensing regime
via:
\be
\bmath{e} = \frac{\bmath{e^s} + \bmath{g}}{1 + \bmath{g^{\star}e^s}}
\ee
from Seitz \& Schneider (1997), where $\bmath{e}$ is a complex variable and
$\bmath{g}$, $\bmath{g^{\star}}$ are the reduced shear and its complex conjugate
respectively. The complex ellipticity is represented in terms of two components 
$\bmath{e}=e_1+i e_2$. In this formalism, we expect that $\langle \bmath{e} \rangle = \bmath{g}$
for an unbiased sample where the average intrinsic ellipticity is zero, 
$\langle \bmath{e^s} \rangle = 0$. As such we will use $\langle \bmath{e} \rangle$ 
for a sample of galaxies as our estimator of shear $\bmath{g}$.
For a population of galaxies we integrate over the probability distribution in 
ellipticity of the sample, $f(\bmath{e})$, to obtain the expectation value of ellipticity 
$
\langle\bmath{e}\rangle = \int \bmath{e} f(\bmath{e}) d\bmath{e}.
$
In the Bayesian formalism we can write a similar expression for an individual galaxy 
if we know its Bayesian posterior probability distribution and hence
 for a sample of $N$ galaxies we can evaluate the sample mean as
\be
\langle\bmath{e}\rangle = \frac{1}{N} \sum_i \int \bmath{e} p_i(\bmath{e}|\bmath{y}_i) d\bmath{e}.
\ee
This allows error estimates to be made
on a galaxy-by-galaxy basis and its contribution to the signal to be evaluated.

In measuring shear we cannot know in advance the correct prior to apply,
even if we know the intrinsic unsheared ellipticity prior distribution, because 
the amount of shear varies over the sky in a way that we are attempting to measure.
We must therefore use a prior that contains zero shear. The result of having 
a zero-shear prior introduces the need to add a weight to the ellipticities to 
counter the effect of this assumption. This 
{\em shear sensitivity} is an effect which has been identified by a number of 
shape measurement methods for example Bernstein \& Jarvis (2002), Luppino \& Kaiser (1997), 
Kaiser (2000) and Massey et al. (2007a)
(it has also been called shear `polarisability' or `responsivity'). 
Crucially the Bayesian methodology allows the 
magnitude of this effect to evaluated on a galaxy-by-galaxy basis directly from the data. 

The shear sensitivity for an individual galaxy may be quantified as 
$|\partial\langle\bmath{e}\rangle_i/\partial\bmath{g}|$: 
a measure of how the measured mean ellipticity
$\langle\bmath{e}\rangle_i$ for the $i^{\rm th}$ galaxy depends on the shear $\bmath{g}$.  
For measurements on noisy data we expect the sensitivity to be reduced from the ideal value of unity.
For a given sample of $N$ galaxies the estimator of the shear is now given by 
\be
\label{key1}
\hat{\bmath{g}} = \frac{\sum_i^N \langle \bmath{e}\rangle_i}
{\sum_i^N |\partial\langle\bmath{e}\rangle_i/\partial\bmath{g}|}.
\ee
This is the key equation used to estimate the shear. 
The shear sensitivity for an individual galaxy should lie in the range 
$0 < \partial\langle\bmath{e}\rangle_i/\partial\bmath{g} \leq 1$, for a measurement completely 
dominated by noise $\partial\langle\bmath{e}\rangle_i/\partial\bmath{g}\sim 0$.

The shear sensitivity can be calculated to first-order using the likelihood and the prior 
probability distributions for an individual galaxy using 
\be
\label{partial}
\frac{\partial\langle\bmath{e}\rangle}{\partial\bmath{g}} \simeq 
1-\frac{
\int \left(\langle\bmath{e}\rangle-\bmath{e}\right) {\mathcal L}(\bmath{e}) \frac{\partial{\mathcal P}}{\partial\bmath{e}} d\bmath{e} 
}
{
\int{\mathcal P}(\bmath{e}){\mathcal L}(\bmath{e})d\bmath{e}.
}
\ee
In the case that 
${\mathcal P}(\bmath{e})$ is fitted with a function 
$\frac{\partial{\mathcal P}}{\partial\bmath{e}}$ can be evaluated analytically. 

The summations over the posterior probabilities, used here to find the mean
ellipticity and hence shear of a sample, could be replaced by a convolution of
all the posterior probability distributions. For a set of $N$ galaxies with
mean ellipticity $\langle\bmath{e}\rangle$ this would yield the probability
distribution $P(N\langle\bmath{e}\rangle)$ whose expectation value is given
by $N$ times the mean: that we calculate here by summation. This would be a
useful procedure for making weak-lensing maps. For cosmological studies we
may be more interested in quantities such as shear variance or the shear
power spectrum, for which the calculation of a full posterior probability
distribution is less straightforward. We leave a full discussion of this
issue for a future publication.

\subsection{Overview of Fast Realistic Galaxy Model Fitting}
\label{Overview of Fast Realistic Galaxy Model Fitting}
The method we use to evaluate the likelihood of a galaxy's ellipticity ${\mathcal L}(\bmath{e})$ 
is to attempt to fit a model surface brightness profile to each galaxy image. 
For a simple model galaxy
whose profile is parameterised by a characteristic radius the total number of free parameters that 
need to be estimated is six: position (two parameters), ellipticity (two parameters), brightness 
and the radius. The key innovation of the work presented in Miller et al. (2007) is that if the model 
fitting is done in Fourier space then the marginalisation over position and brightness can be done 
analytically therefore speeding up the model estimation, leaving only the radius to be marginalised 
over to obtain the ellipticity likelihood ${\mathcal L}(\bmath{e})$. 
By using fast Fourier transform techniques the method can provide a full likelihood surface for an 
individual galaxy in $\sim 1$ second (on a standard $1$ GHz CPU). 

As shown in Miller et al. (2007) the likelihood of a model galaxy being the correct fit 
to a galaxy image can be written as
\be
\label{likee1}
{\mathcal L} \sim \sqrt{\frac{2\pi}{A}} e^{-\sum y_i^2/2\sigma_i^2} e^{AB^2/2}
\ee
this has been analytically marginalised over the amplitude of the model, 
$A$ and $B$ are summations over combinations of the data $y_i$ and model $y^m_i$ defined in 
Miller et al. (2007), $\sigma_i$ is the statistical
uncertainty of the data. 

To marginalise over position it is more straightforward to work in Fourier space where the 
data and model vectors can be rewritten as 
\be
y_i = \sum_k y_k e^{-i\bmath{k.x_i}}, \hspace{1cm}
y^m_i = \sum_k y^m_k e^{-i\bmath{k.x_i}}.
\ee
One can simplify the various summations by assuming that faint
galaxies are being used in weak lensing measurement, such that $\sigma_i$ is dominated by the
background photon shot noise and is constant for all pixels. This assumption of spatially invariant 
noise is applicable to faint galaxies but not for very bright galaxies, but since weak lensing 
is concerned with faint galaxies this assumption is valid. 
To take into account the effect of position uncertainty a 
shift $\bmath{X}$ is introduced into the model position, so that the new model becomes
\be
{y^m_i}' = \sum_k y^m_k e^{-i\bmath{k.x_i}} e^{-i\bmath{k.X}}.
\ee
Substituting into equation (\ref{likee1}) the likelihood becomes
\be
{\mathcal L} \propto 
\exp\left[ 
\frac{|h(\bmath{X})|^2}{2\sigma^2 \sum {y^m_i}^2 } 
\right]
\ee
where $h(\bmath{X})$ is the cross-correlation of the data $y_i$ with the model
$y^m_i$.
To marginalise over $\bmath{X}$ Miller et al. (2007) adopt a prior on position 
chosen to be a Gaussian centred on some previously estimated position 
and that falls off to zero at large distances.

If the cross-correlation function has the Gaussian form $h = h_0 \exp [-(r-r_0)^2/s^2]$
and we approximate the likelihood itself as a Gaussian it can be shown that  
the likelihood, now marginalised over position and amplitude becomes 
\begin{equation}
{\mathcal L} \propto \frac{\pi s^2}{2b^2} \frac{e^\beta }{\beta } e^{-r_0^2/2b^2}
\label{eq:likelihood}
\end{equation}
where $\beta$ depends on the amplitude of the cross-correlation, 
$r_0$ is the nominal galaxy position, $s$ 
is the variance of the cross-correlation and $b$ is the error on the galaxy position.
This is another key equation which is used in the {\sc lensfit} implementation i.e. 
if the width $s$, amplitude $h_0$ and centroid $r_0$ of the cross-correlation
function can be determined then the marginalised likelihood may be estimated 
from equation (\ref{eq:likelihood}). 

For our model we choose the de Vaucouleurs profile which has been shown to be a good estimate of 
realistic galaxy profiles. We justify the choice of profile since 
for faint galaxies exponential and de Vaucouleurs 
profiles are indistinguishable. The STEP simulations present a real challenge of this choice 
since STEP1 consists of composite exponential$+$de Vaucouleurs profiles and STEP2 uses complex 
galaxies morphologies.   
To create a sheared set of galaxy models the axial ratio $\beta$ 
and orientation of the model $\phi$ are related to ellipticity using equation (\ref{axialr}). 
So in the case of a de Vaucouleurs profile equation (\ref{eq:likelihood}) yields a likelihood as
a function of $e_1$, $e_2$ and the scale radius of the model ${\mathcal L}(e_1,e_2,r)$. To obtain 
the likelihood as a function of ellipticity, so that it can be used in equation (\ref{key1}), 
we analytically marginalise over the radius using a simple summation 
\be
{\mathcal L}(e_1,e_2)=\int dr {\mathcal L}(e_1,e_2,r)\approx
\sum_{r_{\rm min}}^{r_{\rm max}} {\mathcal L}(e_1,e_2,r) \Delta r
\ee
where $r_{\rm min}$ and $r_{\rm max}$ are some minimum and maximum that are 
numerically justified in the following Section \ref{Numerical Convergence}.

\subsection{Numerical Convergence}
\label{Numerical Convergence}
An important feature of this shape measurement method is that there are 
no parameters which are tuned
or changed in order to create an unbiased shear estimator. Originally the issue 
of tuning a shape measurement pipeline was raised by Bacon et al. (2001) who found that 
to relate measured shear to input shear a factor of $0.8$ was needed for their 
particular KSB (Kaiser et al., 1995) implementation. 
Some methods, including other KSB implementations, do not require 
tuning parameters. However what existing shape measurement methods find, and KSB somewhat more than 
others, is that there are large magnitude and size dependent biases (Massey et al., 2007) and that 
tuning is required to eliminate these biases to some degree.   
Recently Schrabback et al. (2007) and Leauthaud et al. (2007) find 
that they need ``shear calibration'' factors of up to $0.8$ so that their shear measurement 
pipelines agree with simulations to the $10^{-2}$ level over a range of magnitude.  

We choose a grid in $e_1$, $e_2$ and $r$ to search the parameter space, however 
this is not essential and one could imagine using a Monte
Carlo Markov Chain (MCMC) if this was preferable in terms of speed or accuracy. 
The level of accuracy with which the parameter space will need to be characterised, and hence 
the results of numerical convergence, will 
depend on the data set used particularly on the signal-to-noise of the galaxies. 
For example in the case that the likelihood surfaces were sharp delta 
functions the parameter space may need to have a finer sampling than if the likelihood surfaces 
were broad. Since weak lensing surveys target faint galaxies the likelihood surfaces 
will always be broad so that finite grid sampling should be the fastest method. In the case of the 
STEP simulations we found that the likelihood and prior surfaces were broad enough that they could be 
characterised to sufficient accuracy, such that any parameters of interest numerically converged, 
in fewer steps than an MCMC algorithm could have characterised the probability surfaces. 

\begin{figure}
\resizebox{84mm}{!}{
\includegraphics{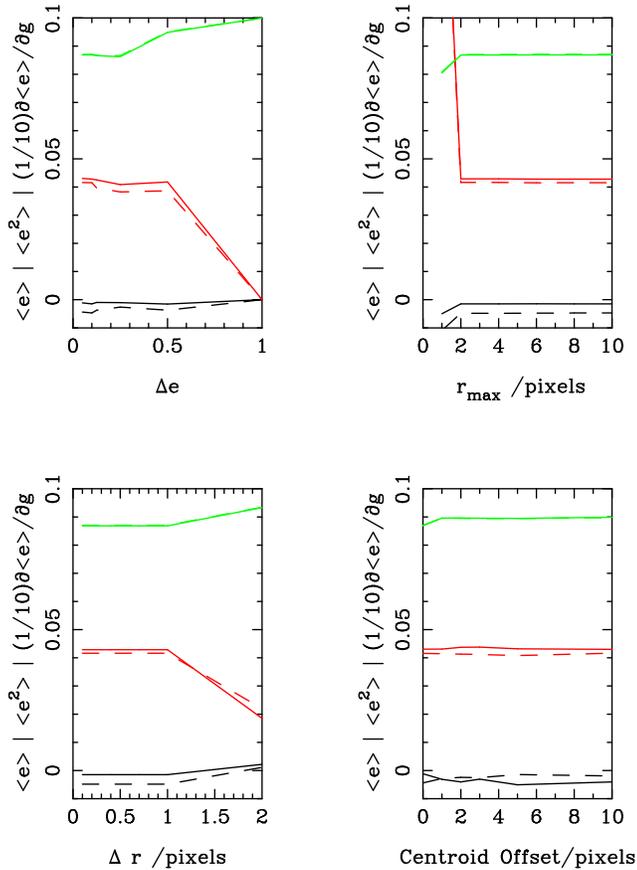}}
\caption{Variation in the expectation value of the ellipticity $\langle e_i\rangle$ 
  (black, lowest lines), the variance in ellipticity $\langle e^2_i\rangle$ 
  (green, middle lines) and the 
  shear sensitivity $d e_i/d g_i$ (red, upper lines) for $\gamma_1$ (solid lines) and $\gamma_2$ 
  (dashed lines) as a function of the numerical values used in the {\sc lensfit} code. 
  Note the y-axis' displays the value of each of these quantities $\langle e_i\rangle$, 
  $\langle e^2_i\rangle$ and $\partial \langle e_i\rangle/\partial g_i$ individually; note that we 
  scale the sensitivity by $1/10$. The top two panels and the 
  lower left panel show how the parameters used to estimate shear vary as 
  the numerical parameters used in {\sc lensfit} are varied; these are 
  the range $r_{\rm max}$ and the resolution $\Delta r$ used to find the likelihood as a function 
  of galaxy radius, and $\Delta e$ the resolution in ellipticity. As each 
  parameter is varied the others are kept at the values of $\Delta e=0.1$, $\Delta r=0.2$ pixels 
  and $r_{\rm max}=10$ pixels, which are the values that we use in the remainder of the paper. 
  The lower right panel shows how the parameters of interest vary as the nominal catalogue 
  position of every galaxy is offset, this tests the ability of the method to marginalise over the 
  position of the galaxy centroid. 
  The simulation
  used was the STEP1 PSF 0 zero-shear image, the results are for the average over the whole 
  galaxy ensemble.}
\label{num_con}
\end{figure}

The only numerical parameters which need to be 
specified are, in the case of a grid search in ($e_1$, $e_2$, $r$),
the resolution in ellipticity $\Delta e$ and the range and 
resolution in the scale factor
$r_{\rm min}$, $r_{\rm max}$ and $\Delta r$. Figure \ref{num_con} shows that the values 
measured from the {\sc lensfit} code, which are of interest in shear estimation, are all convergent 
in a certain regime. 
To ensure that the code is numerically stable we use values of $\Delta e=0.1$, $\Delta r=0.2$ pixels 
and $r_{\rm max}=10$ pixels. For $r_{\rm max}$ and $\Delta r$ these values are well within the 
numerically stable regime, these are chosen so that the code has some built-in redundancy when 
marginalising over the radius so that the results are assured to be robust, 
this was not a great sacrifice in computational speed since the method still operates at $\sim 1$ 
second per galaxy.
The value of $\Delta e=0.1$ is chosen since it is within the numerically stable regime,  
some redundancy could be built-in at the expense of computational time; 
the time to find the full posterior likelihood scales as $1/\Delta e^2$.
We set the minimum radius investigated to $r_{\rm min}=0$, any 
objects for which the most likely value is $r\equiv 0$ we identify as stars and do not use in the average 
shear estimation. This means our 3D parameter space in ($e_1$, $e_2$, $r$) has a maximum of 
less than $20\times 20\times 50$ points for which the likelihood must be evaluated. 

We also present an investigation into the marginalisation over the position 
of the centroid of the galaxy. The {\sc lensfit} method should be robust to inaccuracies in the 
centroid position of any galaxy since the method analytically marginalises over the center of the 
galaxy model. In Figure \ref{num_con} we introduce a constant 
offset in the position of every galaxy from the actual galaxy position. It can be seen that the 
method is relatively insensitive up to a constant offset of $\sim 10$ pixels 
so that when estimating the position of 
galaxies any source extraction routine could misplace galaxies by up to this amount with no major 
effect on the shear estimation. In reality source extraction routines such as {\sc SExtractor} 
(Bertin \& Arnouts, 1996) or 
{\sc hfindpeaks} (part of the {\sc imcat}\footnote{www.ifa.hawaii.edu/$\sim$kaiser/imcat/content.html}
software package) have accuracies much better than this (Heymans et al., 2006). 

In the implementation of the method we extract a small postage-stamp image about each galaxy. 
This postage stamp size then determines the size of the model and PSF. 
We use a postage stamp size of $32\times 32$
pixels, which is the same for the model, PSF and galaxy images.  
We reject any galaxies which
are in a close pair i.e. ones which have one or more other galaxies within their postage 
stamp. It is possible to intelligently reject close galaxy pairs based on signal-to-noise criterion, 
Schrabback et al. (2007) and Leauthaud et al. (2007) for example employ more sophisticated 
close-pair rejection algorithms however we have not implemented these here. 
The postage stamp size was optimised for the 
STEP simulations; if the postage stamp size is too large then too many galaxies will have `close 
neighbours' (i.e. another galaxy or star in the postage stamp) and be rejected, if too small then the 
largest galaxies will not fit into the postage stamp. We found that a $32\times 32$ stamp was the 
smallest stamp in which every galaxy could fit, allowing for a factor of $2$ to minimise edge effects.
Note the size of the postage stamp does not determine $r_{\rm max}$, and as we have shown a 
value $r_{\max}=10$ pixel is sufficient to ensure numerical convergence.
 
\subsection{PSF Estimation}
\label{PSF Estimation}
The level of accuracy with which a PSF can be characterised is an
important factor in the performance of any shape measurement method. As
described by Massey et al. (2007), the PSF must either be deconvolved
from the image to generate a raw galaxy image or, more robustly, in
{\sc lensfit} a galaxy model is convolved with the PSF and then fitted
to the data. Existing methods usually either stack star images or fit
functional forms to star images. A limitation of all methods is that 
the spatial and chronological variability of the PSF needs to be
determined, for which only a finite number of stars in each image
are available (e.g. Paulin-Henriksson et al., 2007).

We create the PSF model by stacking star images.  The data for each
star are sub-sampled onto a 50-times finer pixel grid using sinc
function interpolation (which precisely preserves the data values
without inventing any new Fourier modes), and stacking takes place in
a two-stage iterative process.  In the first stage, the stars are
coaligned by cross-correlating with a delta function, and then
coadded.  Then, each star is individually compared with the stack by
cross-correlation, any that have a low cross-correlation amplitude are
rejected.  In the second stage, the remaining stars are again
cross-correlated with the stacked PSF to redetermine their centroids
more accurately, the stack is remade and again individual stars are
checked by cross-correlation with the new stack, and eliminated if
appropriate.  The stack of surviving stars thus forms the final PSF
which is then downsampled to the original pixel sampling (without
aliasing since in the above process there have been no modes created
above the Nyquist frequency).  We find in the STEP simulations that if
stars of low signal-to-noise are used, many are rejected at the
cross-correlation stage.  Stars with peak signal-to-noise ratio
greater than 30 worked well, with only a small number of stars being
rejected, these being instances of closely neighbouring stars being
blended together.  No selection of stars ``by eye'' was required; the stars 
were selected using the {\sc SExtractor} `class\_star' parameter.

If the true PSF is band-limited at the pixel sampling Nyquist frequency the
above method produces a faithful representation of it in the sampled
image plane.  Convolution with a galaxy model then yields an
``observational model'' of the galaxy with the effects of the PSF and
pixel sampling correctly matched to the data.  In reality, the
band-limited assumption is not likely to be true, and all methods of
PSF determination and hence galaxy shape measurement are ultimately
limited by the problem of pixelisation: we have no information on the
PSF below the pixel scale, and any Fourier modes in the PSF with
frequencies higher than the Nyquist frequency become aliased to lower
measured frequencies.  There is in principle some information
available on the sub-pixel scale owing to the centres of the stars not
being exactly centred on pixels, but in reality it is very hard to
extract that information to yield a robust estimate of the high
frequency modes in the presence of noise.  Without such information
the best we can do is to assume that the sampling of the PSF is
sufficient to render aliasing of high frequency modes
insignificant. This deficiency of information may become one of the
main limiting factors in the accuracy with which weak lensing shear
may be measured.  Dithering of images would also allow us to gain
back information on the sub-pixel scale and for some future space-based
experiments such as DUNE or SNAP, high resolution pre-launch
characterisation of the PSF should allow improved PSFs to be
reconstructed. Jarvis \& Jain (2005) and Jee et al. (2008) discuss the 
characterisation of a PSF using PCA techniques which can be used for ground-based 
surveys. 

\section{Estimation of the Prior}
\label{Estimation of the Prior}
A requirement of the Bayesian shape measurement approach is the accurate and correct
estimation of the ellipticity prior. Here we present an iterative method that should yield
the correct prior from the data itself (this is similar to the approach introduced 
in Lucy, 1974; Richardson, 1972 and Lucy, 1994 in image deconvolution). 
One could use the entire 
data set or a subset of a large 
wide field survey to do this, many planned future surveys, for example DUNE, Pan-STARRS, SNAP and 
LSST include in their strategies medium-deep surveys over much smaller areas than the main 
wide field surveys which would be used for cosmic shear analysis. These medium-deep 
surveys would be ideal data sets from which to estimate the prior in this fashion. 

As already discussed in Miller et al. (2007) one must assume a prior with zero shear i.e. 
centred on $e_1=e_2=0$, since this is the baseline assumption which enforces no 
\emph{a priori} knowledge on the result. Also, in the case of real data one would expect 
the shear to average to zero over a sufficiently large number of galaxies. In the STEP 
simulations there is a large shear $\gamma\sim 0.05$ to $0.1$ over a whole image which in 
reality one would not expect, the simulations thus test this assumption of a zero-centred 
prior to an extreme.  
When testing on simulations the prior has to be found using a zero-shear image since the 
posterior probability estimated from these images 
will be the intrinsic ellipticity distribution, however in 
a real data set where the mean shear across an image should be zero the prior can 
be estimated directly from the data. 
Note again that the level of bias introduced by this assumption can be
exactly accounted for within the Bayesian formalism by using the shear sensitivity, equation
(\ref{key1}). 

The iterative approach centres around equation (\ref{cent}) which is the average summed 
posterior probability for an ensemble of $N$ galaxies
\be
\langle
\frac{1}{N}
\sum_{\alpha} {\mathcal P}(\bmath{e}){\mathcal L}(\bmath{e})_{\alpha}
\rangle= {\mathcal P}(\bmath{e}).
\ee
If the prior which was initially used was the true, intrinsic, prior this is a 
stable equation in the case that a sufficient number of galaxies are used 
i.e. if the prior which is output on the right hand side (RHS) of the equation is 
used on the left hand side (LHS) 
of the equation in a second iteration the result will be the same. If the
prior used on the LHS is not the true prior then the distribution given on the RHS will 
be closer to the true intrinsic distribution than the prior initially used. The 
method involves using this equation to iterate on the prior used i.e. 
\be
{\mathcal P}(\bmath{e})_{i+1}=\langle\frac{1}{N}
\sum_{\alpha} {\mathcal P}_i(\bmath{e}){\mathcal L}(\bmath{e})_{\alpha} 
\rangle
\ee
this is repeated over $i$ iterations, when the prior used is an accurate and correct 
representation of the true intrinsic prior a stable solution will have been found. Note that
the prior ${\mathcal P}_i(\bmath{e})$ is normalised; and as such ${\mathcal P}_i(\bmath{e})=\sum 0=0$ 
is not a stable solution. This iterative approach, and the method in general, 
assumes that the function that is 
used as the prior is differentiable and non-zero at all points in the parameter space at which 
the likelihood is evaluated. 

Crucially the usual concerns involved with iterating on a data set do not apply here. This 
is due to the nature of the operation we are using. We are
not using a prior to improve the probability distribution of some estimated parameter, but 
rather using the data to estimate the prior.  
Since the operation described above yields the prior itself once the true prior is found 
this operation could be performed given a sufficient number of galaxies 
\emph{ad infinitum} with no divergence of results. In the limit of a small number of galaxies 
this stability will diverge due to shot noise (i.e. sampling variance) in ellipticity, in 
Section \ref{Testing the Iterative Approach with STEP} 
we estimate the minimum sample sizes that are needed for convergence to the 
correct prior.

\subsection{Fitting the Prior}
\label{Fitting the Prior}
In practice after each iteration we fit the prior surface with a functional form and use 
this as the prior for the next iteration. This is done since a functional form ensures 
that the prior is smooth and known everywhere, also using a functional form means that the 
derivative of the prior, to use in equation (\ref{partial}) 
can be calculated exactly. Since the prior must be differentiable and non-zero 
we do not allow the functional form to have turning points in the region $0 < |e| <1$. 
Note this ensures that no stopping criterion is needed since by 
fitting a functional form any noise in the probability distribution is averaged over (the smoothing acts like 
a regularising constraint), if a 
functional form were not used then the iterations could artificially amplify any noisy structures (this 
is a concern in using the iterative approach in image deconvolution, Richardson, 1972; Lucy, 1974). 
The 2D functional form in ($e_1$, $e_2$) we use is
\be
\label{funcprior}
{\mathcal P}(e_1, e_2)=A\cos\left(\frac{|e|\pi}{2}\right)
\exp\left[-\left(\frac{2|e|}{B(1+|e|^D)}\right)^C\right]
\ee
where $B$, $C$ and $D$ are free parameters to be fitted and $|e|=\sqrt{e_1^2+e_2^2}$.
The prior is always normalised so that the parameter $A$ is determined by the normalisation.  
The cosine factor ensures that the prior goes to zero at $|e|=1$. We have found this to be 
a good fit to both the STEP simulation's 
intrinsic distributions and the APM survey's published
intrinsic ellipticity distributions (Crittenden et al., 2001). 
To convert to a 1D distribution in $|e|$ one must 
multiply by the appropriate parameter space volume factor i.e. 
${\mathcal P}_{\rm 1D}(|e|)=2\pi|e|{\mathcal P}(e_1, e_2)$. To fit the output prior from 
each iteration with this 2D functional form we minimise the 
\emph{cross-entropy} defined as  
\be 
H(p,q)=-\sum_x p(x)\log q(x)
\ee
where $q(x)$ is some estimated probability distribution and $p(x)$ is the `true' distribution.  
This is similar to the Kullback-Leibler divergence between two distributions and is 
a measure of the difference between the two distributions $q(x)$ 
and $p(x)$. Formally it
measures the average number of bits needed to identify an event from a set of 
possibilities 
if the probability distribution $q$ is used, rather than the true distribution $p$. In our case
we wish to minimise the difference between the functional prior and the output prior 
\be 
H=-\sum_{e_1}\sum_{e_2} {\mathcal P}(e_1, e_2)_{\rm functional \ form}\log {\mathcal P}(e_1, e_2)_{\rm output}. 
\ee
By minimising this function the best fit functional form to the output prior is found. 
We found this to be more robust and yields better fits 
to the STEP intrinsic ellipticity distributions than projecting the distribution onto a 
1D function of $|e|$ and using a binned least squares fitting method.

\subsection{Testing the Iterative Approach with STEP}
\label{Testing the Iterative Approach with STEP}
To test this iterative method we estimated the prior of the STEP1 simulations for PSF 0 from 
the zero-sheared image (PSF 0, image $0$; see Section \ref{Application to the STEP1 Simulations} 
for a full description of the STEP1 simulation) and compared the prior found with the input  
intrinsic ellipticity distributions used to create the simulated images. 
\begin{figure}
\resizebox{84mm}{!}{
\includegraphics{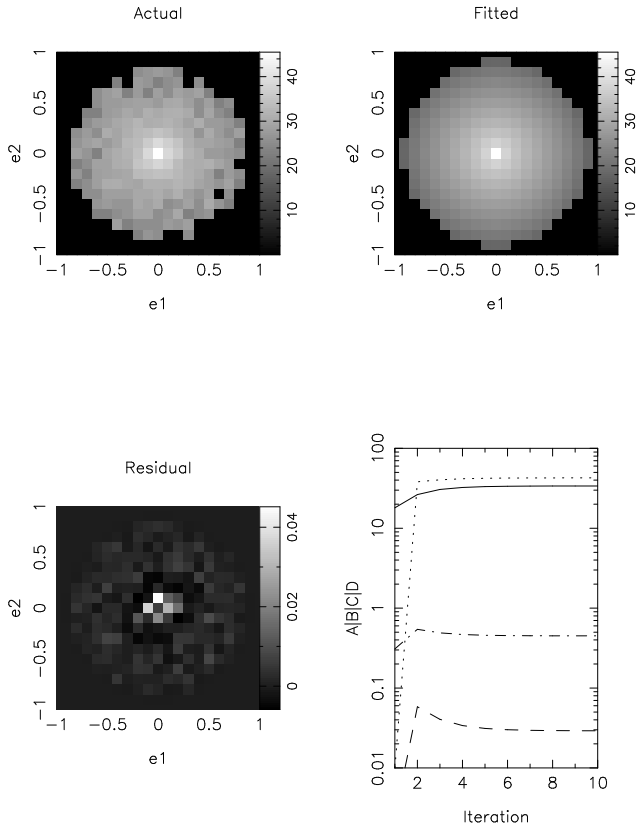}}
\caption{The top panels show the actual intrinsic ellipticity distribution used in the STEP1
simulations and the prior created using the iterative method, 
the distributions are normalised. The left hand side of the lower panel shows 
the residual as a fraction of the actual distribution i.e. 
$R=\Delta P/P=(P_{\rm true}-P_{\rm func})/P_{\rm true}$.
The lower righthand panel shows how the values of 
the parameters of the fitted functional form of the prior change as the number of iterations
increases for parameter A (solid line), B (dashed line), C (dot-dashed line) and D (dotted line), 
see equation (\ref{funcprior}).}
\label{priorexample}
\end{figure}
Figure \ref{priorexample} shows the actual intrinsic ellipticity distribution in 
$(e_1, e_2)$ for the STEP1 simulation and the prior found using the iterative approach. 
It can be seen from the very low level of fractional residual 
$\Delta P/P=(P_{\rm true}-P_{\rm func})/P_{\rm true}$ between the ``true'' and the estimated 
prior, of order $0.02$, that the iterative approach is an accurate and good method for 
finding the correct prior. Furthermore the convergence to an accurate functional fit can 
occur in approximately $5$ iterations. We tested the robustness of this convergence to the 
starting values of the functional parameters ($A$ -- $D$) and found that in all cases there 
was convergence in fewer than $6$ iterations.  

The correct prior is formally only a stable solution to the iterative approach 
in the case of an infinite ensemble of galaxies. 
Here we present results that show the variation of the estimated prior 
as a function of the number of 
galaxies used in the iterative determination. There is no simple analytical way to determine 
the minimum number of galaxies required to determine the prior to a certain level of accuracy 
as this depends on the form of the prior. 
To accurately determine the prior probability surface the 
ellipticities of the galaxies used have to sample to some degree the whole ($e_1$, $e_2$) plane 
i.e. if a subset of galaxies were used that had exactly the same ellipticity they would not recreate 
the intrinsic distribution of the overall population using the iterative approach. In the limit of a 
small number of galaxies, from which the ellipticity is imperfectly determined, sample shot noise 
will become an important factor. The accurate determination 
of the prior from a subset of galaxies from a population thus depends in a complex way on 
the shape of the likelihood surfaces from those galaxies and the number used. One may expect that 
$\sim 100$ galaxies would not suffice since, with a resolution of $\Delta e=0.1$, we 
evaluate the prior at $\ls 100$ independent points in the ($e_1$, $e_2$) plane. 

\begin{figure}
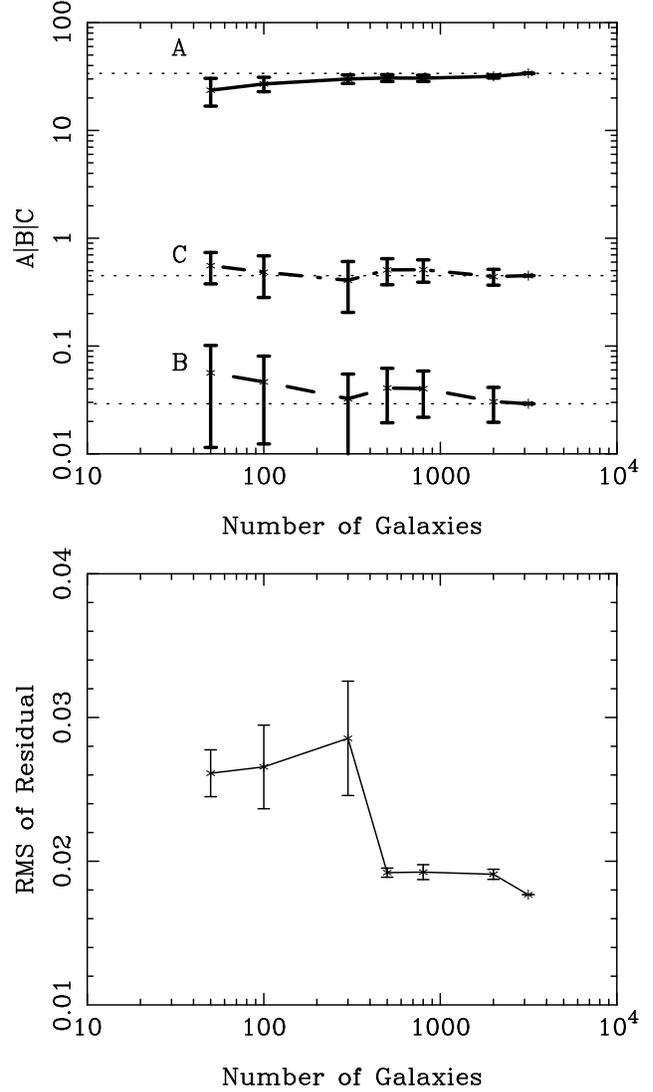

\resizebox{84mm}{!}{
\includegraphics{step1_numgal_prior.ps}}
\resizebox{84mm}{!}{
\includegraphics{rms_prior_residual.ps}}
\caption{The top panel shows the variation in the parameter values, from equation (\ref{funcprior}), 
  found by recovering the prior using the iterative approach (after $10$
  iterations) as a function of the number of galaxies used in the prior estimation. 
  For each galaxy number bin 
  we selected ten random sub-populations of the entire sample of galaxies. The bold 
  black lines show the mean and the variance of the parameter values over all 
  the random samplings. 
  We show parameters A=solid line, B=dashed line, C=dot-dashed line, the dotted horizontal 
  lines show the values of these parameters when the whole galaxy sample is used. 
  We do not show the variation of D for clarity, and since the value of D has a small effect on 
  the functional form of the probability; the value of D begins to deviate at $\sim 500$ galaxies.
  The bottom panel shows how the rms of the fractional residual 
  between the actual STEP1 input prior and 
  the functional fit to the prior, i.e. 
  $R=\Delta P/P=(P_{\rm true}-P_{\rm func})/P_{\rm true}$, 
  varies with the number of galaxies used to create the 
  prior. The mean is the average rms over all random samplings of the full input catalogue, the 
  error on each point shows the variance of the rms over the random samplings. 
}
\label{step1_numgal_prior}
\end{figure}

We numerically 
investigated the number of galaxies required to estimate the prior 
by selecting random samples of galaxies from the STEP1 PSF0 
catalogue and recreating the prior using only these galaxies for many different random realisations 
of the sub-set. The prior created using these random sub-sets 
was then be compared to the prior found using the entire population. In 
Figure \ref{step1_numgal_prior} we show the best fit values of the functional parameters 
as the number of galaxies used to estimate the prior changes. For each galaxy sub-set number we made
$10$ random samplings of the full catalogue, the lines show the mean values of the 
functional parameters averaged over these random samplings (after $10$ iterations of the prior 
estimation algorithm). The error bars show the the variance in the values over the random samplings 
of the catalogue. 
This shows how the parameters fitted to the prior vary with the number of galaxies 
used to create the prior. The value of the D parameter begins to deviate at a very low level when 
$\ls 500$ galaxies are used, however this parameter has a very small effect on the functional 
form. At $|e|\sim 0$ the D parameter only enters as a second order term, and at $|e|\sim 1$ 
the cosine factor dominates which suppresses any influence that this parameter may have had. 

The deviation and variance in parameters A, B and C 
becomes significant when $\ls 100$ galaxies are used 
i.e. the parameters which fit the prior depend strongly on the specific sub-set of 
galaxies which are randomly chosen, as is 
demonstrated by the variance in the best fit values increasing in the top panel of 
Figure \ref{step1_numgal_prior} when fewer galaxies are used, 
in addition the mean values deviate from the
parameter values found using the whole ($\sim 3000$) population by a large amount.  
The bottom panel of Figure \ref{step1_numgal_prior} shows the root-mean-square (rms) value of the 
residual between the actual STEP1 input prior and the functional fit to the prior as a function 
of the number of galaxies used in the iterative approach. It is clear that the rms of the 
residual increases 
dramatically when the number of galaxies falls below $\sim 300$. 

When analysing the STEP simulations, in which there are a small number of galaxies per image, 
the problem of too few galaxies with which to recover the prior will be encountered. This is 
discussed in Section \ref{Application to the STEP2 Simulations} where we find that in the 
STEP2 simulations the intrinsic ellipticity varies as a function of size and magnitude, and that 
by correctly accounting for this variation the shear estimation can improve. 
In an actual survey in which the number of galaxies is $\gg 10^4$ 
one would expect that in any sub-population of galaxies, defined by some commonly  
observed property for example magnitude, size, colour or type  
there would be $\gg 100$ galaxies so that this 
problem will not arise when this shape measurement method is used on data sets. 

\subsection{Summary of the lensfit shape measurement method}
\label{Summary of the lensfit shape measurement method}
Before presenting the results of using {\sc lensfit} on simulations we summarise the 
method. To summarise and clarify we consider the 
$i^{\rm th}$ shear component $g_i$ where $g=g_1+ig_2$ and $\bmath{e}=e_1+ie_2$. We also 
recast any integrals as summations, as is done in the actual {\sc lensfit} implementation. 

\begin{enumerate}[i)]
\item
We use a Bayesian estimator of shear which is given by the summation over 
$N$ galaxies 
\be
\label{key2}
\hat{\bmath{g_i}} = \frac{\sum_{\alpha}^N \langle \bmath{e_i}\rangle_{\alpha}}
    {\sum_{\alpha}^N |\partial\langle\bmath{e_i}\rangle_{\alpha}/\partial\bmath{g_i}|}.
\ee
where $\partial\langle\bmath{e}\rangle_{\alpha}/\partial\bmath{g}$ is the \emph{shear sensitivity}. 
\item
The expectation value of the $i^{\rm th}$ ellipticity value for an individual galaxy $\alpha$ 
$\langle \bmath{e_i}\rangle_{\alpha}$ is given by 
\be
\label{expecte}
\langle \bmath{e_i}\rangle_{\alpha}=\int de_j \int de_i e_i p_{\alpha}(e_i,e_j)\approx 
\sum_j \sum_i \Delta e^2 e_i p_{\alpha}(e_i,e_j)
\ee
where $p_{\alpha}(e_i,e_j)={\mathcal P}(e_i,e_j){\mathcal L}_{\alpha}(e_i,e_j)$ 
is the posterior ellipticity probability distribution for a given galaxy. 
\item
The shear sensitivity is recast from equation (\ref{partial}) as  
\be
\frac{\partial\langle\bmath{e_i}\rangle}{\partial\bmath{g_i}} \simeq 
1-\frac{
\sum_j \sum_i \Delta e^2\left(\langle\bmath{e_i}\rangle-\bmath{e_i}\right) {\mathcal L}(e_i,e_j)
\frac{\partial{\mathcal P}(e_i,e_j)}{\partial\bmath{e_i}}
}
{
\sum_j \sum_i \Delta e^2{\mathcal P}(e_i,e_j){\mathcal L}(e_i,e_j)
}.
\ee
\item 
To calculate the likelihood as a function of ellipticity we use a model fitting approach 
that marginalises over position and amplitude in an analytic way and fits a 
de Vaucouleurs profile. Using equation (\ref{eq:likelihood}) the likelihood is then given as 
a function of radius $r$ and ellipticity $e_1$ and $e_2$. This is then analytically marginalised 
over radius using 
\be
{\mathcal L}(e_1,e_2)\approx\sum_{r_{\rm min}}^{r_{\rm max}} {\mathcal L}(e_1,e_2,r) \Delta r, 
\ee
where we assume a uniform prior in $r$.
\item 
The prior ${\mathcal P}(e_i,e_j)$ is a zero-centred function which is representative 
of the intrinsic ellipticity distribution. We calculate this using an stable iterative approach 
in which the data itself can be used to estimate the prior.  
\end{enumerate}

\section{Results of tests on simulations}
\label{Results of tests on simulations}
In the following Section we describe the simulations in detail and 
present the results of recovering the input shear from these
simulations using {\sc lensfit}, we also compare with the currently published STEP1 and STEP2 
results. 

The ability of a shape measurement method to recover the input shear from 
a simulation in the STEP papers is parameterised by
\be 
\label{M_c}
\gamma^M_i-\gamma^T_i= m_i\gamma^T_i + c_i
\ee
where $\gamma^T_i$ is the `true' (input) shear for the $i^{\rm th}$ shear component and 
$\gamma^M_i$ is the `measured' or estimated shear value using a given shape measurement method. 
$m_i$ characterises any bias in a shape measurement method, $c_i$ 
characterises any residual shear offset. 
Any residual shear offset is usually due to inaccuracies in the PSF estimation, 
a PSF which is slightly more elliptical than reality will simply act to add a constant 
to any estimated shear value. In STEP1 some methods also require a quadratic term on the LHS side of 
equation (\ref{M_c}) $q(\gamma^T_i)^2$, we find that this 
extra term is not required to model our results and present results in terms of 
$m_i$ and $c_i$ in line with the STEP papers.  

\subsection{Application to the STEP1 Simulations}
\label{Application to the STEP1 Simulations}
STEP1 (Heymans et al., 2006) uses simulated galaxies 
which consist of a de Vaucouleurs bulge plus an exponential disk. The simulations are provided with 
six different PSFs (named $0$ to $5$), for each PSF there are $5$ shear sets each consisting of an 
ensemble of $64$ individual images $4096\times 4096$ pixels. 
The $5$ shear sets for each PSF have different shear values of 
$\gamma_1=0.0$, $0.005$, $0.01$, $0.05$ and $0.10$; $\gamma_2=0.0$ is set for all the STEP1 
simulations. 
Each of the shear sets ($64$ images) contains $\sim 2\times 10^5$ galaxies.  
Each image also contains $\sim 3000$ stars from which the PSF can be
determined, the pixel scale in the simulations is $0.206$ arcseconds and the average 
PSF FWHM is $0.8$ -- $0.9$ arcseconds. 

In order to test the {\sc lensfit} method we used {\sc SExtractor} (Bertin \& Arnouts, 1996) 
to create input catalogues from the STEP1 
simulations to find the positions of the galaxies and stars. 
For each image in each PSF set and for each shear value we recalculated the PSF from the stars 
available in that image. We calculated the prior, as described in 
Section \ref{Estimation of the Prior}, using the zero-shear image from each PSF set. 

For the error on the shear estimate $\gamma_i$ and hence $\gamma^M_i-\gamma^T_i$
for each image we use the error on the mean ellipticity given for 
$N$ galaxies by $\sigma_M=\sigma/\sqrt{N}$. 
The expectation value 
$\sigma^2=\langle e_i^2\rangle/(\partial\langle\bmath{e}\rangle_i/\partial\bmath{g})$ 
is calculated by integrating over 
the posterior probability as in equation (\ref{expecte}). We then use a $\chi^2$ fit to 
$\gamma^M-\gamma^T$ as a function of $\gamma^T$ to find the best fit values of $m_i$ and $c_i$ 
defined 
in equation (\ref{M_c}). The errors on $m_i$ and $c_i$ are found by exploring the whole 
($m_i$, $c_i$) 
parameter space and projecting the two parameter $1$-$\sigma$ errors onto the corresponding axis 
to find the $1$-$\sigma$ error on each parameter. 

\begin{figure}
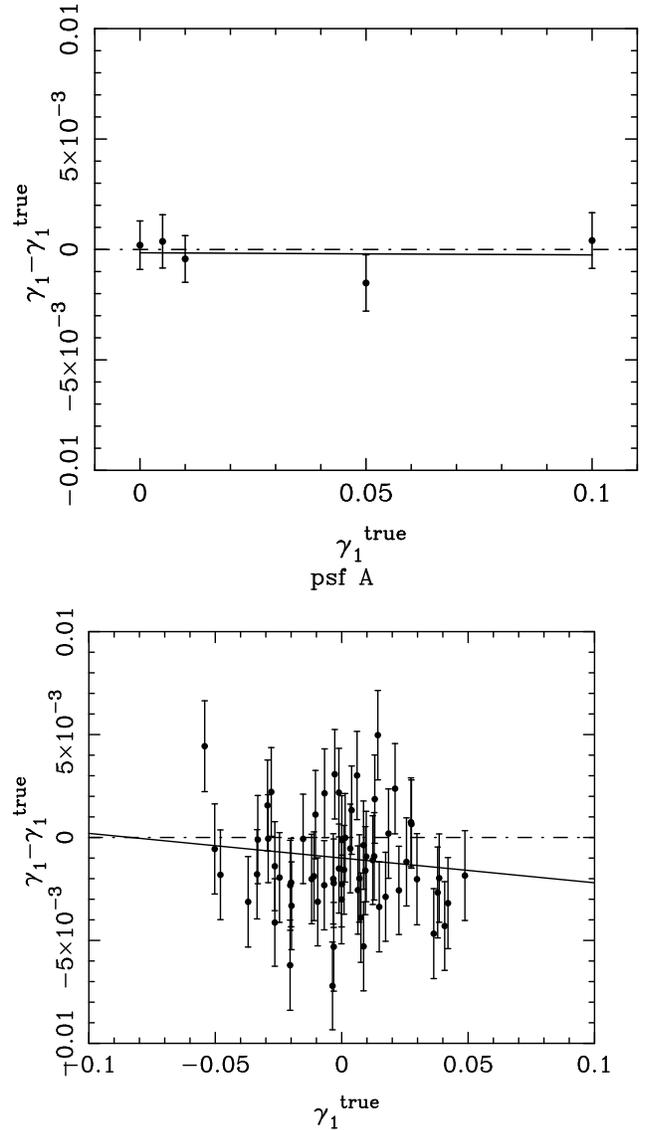

  \resizebox{84mm}{!}{
    \includegraphics{step1_gamma.ps}}

  \resizebox{84mm}{!}{
    \includegraphics{step2_gamma.ps}}

 \caption{The upper panel shows the estimated $\gamma_1$ shear values 
   minus the true (input) $\gamma^{\rm true}_1$  shear for STEP1 PSF 0, note
   that for STEP1 only $5$ input shear values are provided. 
   The upper solid line shows the $m_1$ and $c_1$ fit for STEP1 PSF 0 
   ($m_1=-0.0009$, $c_1=-0.0002$). 
   The lower panel showes the estimated $\gamma_1$ shear values
   minus the true (input) $\gamma^{\rm true}_1$ shear for STEP2 PSF A simulations, note that 
   STEP2 provides $64$ random shear values distributed within the range
   $-0.06 \leq \gamma_1 \leq 0.06$. 
   The lower solid line shows the $m_1$ and $c_1$ fit for STEP 2 PSF A ($m_1=-0.012$, $c_1=-0.00099$).
   Note
   that there are $64$ images which are used to estimate the shear for each point in the upper panel 
   whereas only $2$ images per point are used in the lower panel. 
 }
 \label{gammas}
\end{figure}

Figure \ref{gammas} shows the measured shear minus the true (input) shear $\gamma^M-\gamma^T$ 
for PSF 0 of the STEP1 simulations with the best fit linear function from equation (\ref{M_c}). 

\begin{figure}
  \resizebox{84mm}{!}{
    \includegraphics{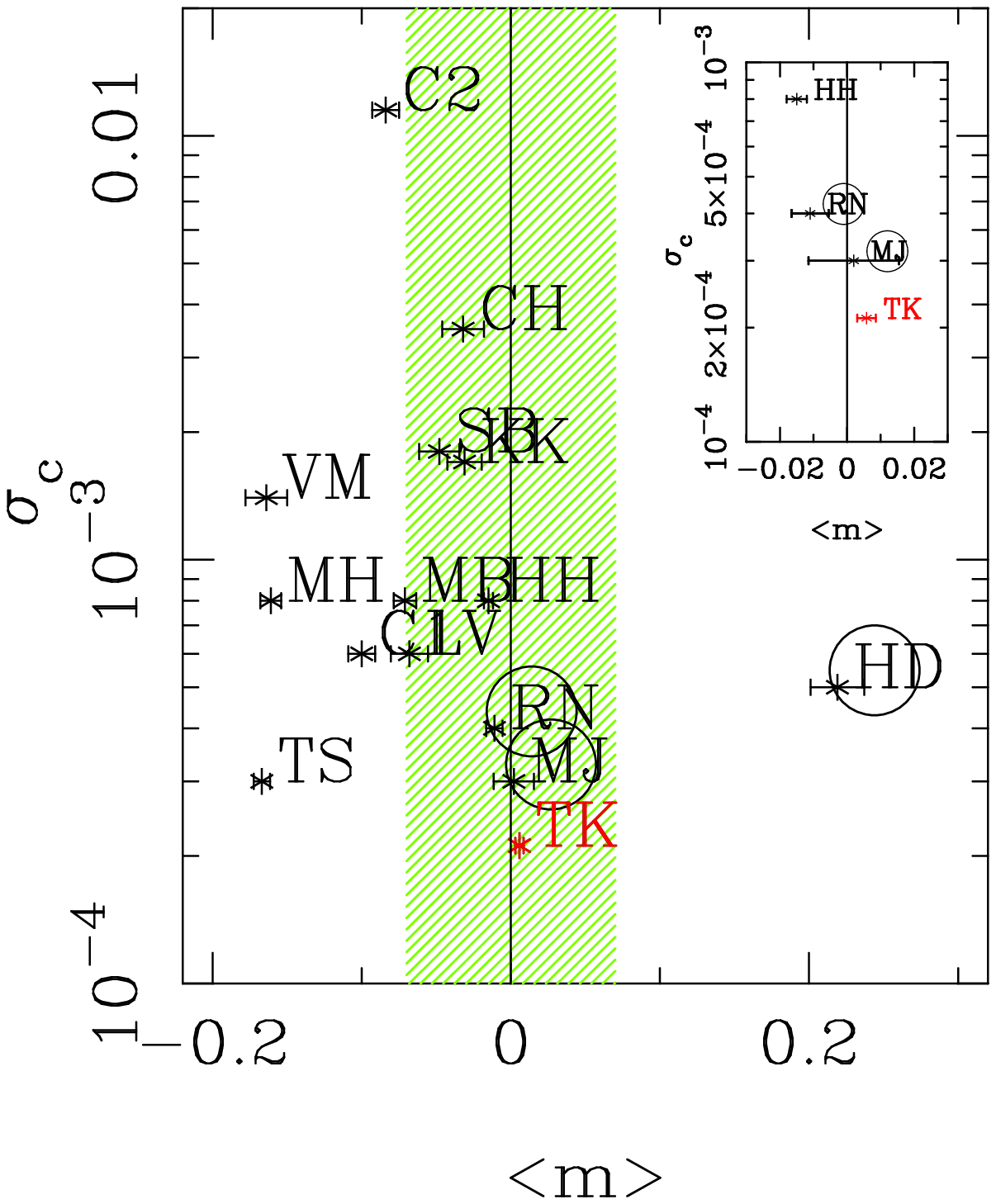}}
 \caption{Adapted from Heymans et al. (2006), Figure 3. The average value of the bias 
   $m$ over all PSFs in the STEP1 simulation and the variation in the offset $\sigma_c$. 
   The red point `TK' shows the result of using {\sc lensfit}, the black points show the 
   other shape measurement methods presented in Heymans et al. (2006) (the labelling reflects the 
   authorship of the method, see Table 2 of Heymans et al., 2006 for more details). 
   The points surrounded by circles are those which required 
   an extra quadratic term in equation (\ref{M_c}). The hatched region indicates 
   the level of precision required by current surveys ($\langle m\rangle\leq 0.07$), 
   as discussed in Heymans et al. (2006). The inset shows a zoom in of the bottom part of the Figure. 
 }
 \label{step1_final}
\end{figure}

Figure \ref{step1_final} shows the results of applying the {\sc lensfit} method to 
the STEP1 simulation. For STEP1 we find $m_1$ and $c_1$, we also find $c_2$ assuming that $m_2=0$, 
as is done in the STEP1 publication. 
$\langle m\rangle$ is the average bias over all PSFs, the error on this value 
is the sum of the squares of the errors on $m$ from each PSF. $\sigma_c$ is the average variance in 
the offset from $c_1$ and $c_2$ i.e. $\sigma_c=\sqrt{\sigma^2_{c1}+\sigma^2_{c2}}$. The 
result is detailed in Table \ref{STEP2table}.
The value of $\langle m\rangle=+0.006\pm 0.005$ 
is the smallest for any method for which a linear fit to 
$\gamma^M_i-\gamma^T_i$ is required (Heymans et al., 2006). 
The methods which require a non-linear term in equation 
(\ref{M_c}), $q(\gamma^T_i)^2$, are shown by a circle about the point in Figure \ref{step1_final}. 
$q>1.3$ for all these methods and, as shown in Heymans et al. (2006) Figure 2, this parameterises 
large non-linear effects. 
The value of $\sigma_c=0.0002$ is smaller than any method in the STEP1 publication. 
This value parameterises any PSF systematics: in 
the absence of systematics and shot noise for a perfect shape measurement method one would 
expect $\sigma_c=0$.  


\subsection{Application to the STEP2 Simulations}
\label{Application to the STEP2 Simulations}

This Section presents the results of extracting shear estimates 
using {\sc lensfit} from the STEP2 (Massey et al., 2007) simulations. These simulations 
consist of shapelet based (Refregier et al., 2003; Massey et al., 2004; 
Massey \& Refregier, 2005) and exponential galaxy profiles convolved with various different PSFs. 
There are six sets of PSF and galaxy profile 
combinations provided. Sets A, C, D, E and F have shapelet simulated galaxies and various 
different PSF shapes. Sets D and E have highly elliptical PSFs aligned along the $e_1$ and $e_2$ 
directions respectively. Set B has the same PSF shape as A but with exponential 
galaxies as opposed 
to shapelet galaxies. Each set consists of $64$ images and $64$ `rotated' 
images. The rotated images 
are exactly the same as the `original' images except that they have been rotated by 
$90^{\circ}$ before being sheared. As
described in Massey et al. (2007) this allows the intrinsic shape noise to be dramatically reduced 
by co-adding the shear estimates from the matching corresponding images. The signal-to-noise 
error on the intrinsic ellipticity is usually given for a sample of $N$ galaxies as 
(equation 3, Massey et al., 2007)
\be 
\langle e^{\rm int}\rangle \approx 0 \pm \sqrt{\frac{\langle (e_i^{\rm int})^2\rangle}{N}}.
\ee
Massey et al. (2007) showed that by defining the average shear as the 
average of the observed ellipticities from the rotated and unrotated galaxy images, 
$\tilde\gamma=(e^{\rm obs,unrot}+e^{\rm obs,rot})/2$, the shot noise error on the average 
shear is reduced to (equation 6, Massey et al., 2007)
\be
\gamma\langle(e^{\rm int})^2\rangle=0\pm \gamma\sqrt{\frac{\langle (e_i^{\rm int})^4\rangle}{2N}}.
\ee
In STEP2 the averaging is done on a galaxy-by-galaxy basis i.e. each galaxy paired with its 
rotated counter-part. 
For the STEP2 simulations $\sqrt{\langle (e_i^{\rm int})^4\rangle}\sim 0.05$, $\gamma<0.06$ and 
$N\sim 1500$ so that the shot noise error on the shear estimate for a given image should be $\sim 
6\times 10^{-5}$ reduced from $\sim 3\times 10^{-4}$. We calculate the shear by taking the mean 
expected ellipticity weighted by the shear sensitivity 
$\tilde\gamma=(\langle e^{\rm obs,unrot}\rangle+\langle e^{\rm obs,rot}\rangle)/2$, 
note that this gives equal weight to the unrotated and rotated probability surfaces. 

Each image (and corresponding 
rotated image) contains $\sim 1500$ galaxies which are usable for shear 
(the images actually contain $\sim 5000$ galaxies 
but the majority are too faint to be detectable), and 
has a different random shear, $\gamma_1$ and $\gamma_2$, applied. The shear values are randomly 
chosen in the range $\gamma\leq 0.06$. For each set (PSF) a star field is provided which contains 
$\sim 240$ stars (and no galaxies) 
which can be used to estimate the PSF, the galaxy fields also contain stars which 
can be used instead of, or in supplement to, the stars provided in the star fields.  
The simulations are a sophistication of the STEP1 simulations in two important ways. 
Firstly the galaxies are
``more realistic'', that is they are mostly shapelet galaxies which exhibit substructure, 
spiral arms etc. This should be a significant test for {\sc lensfit} which assumes 
de Vaucouleurs profiles. Secondly the shear values are varied randomly 
in both the $\gamma_1$ and $\gamma_2$ directions as opposed to sampling 
just $5$ points in $\gamma_1$ and setting $\gamma_2$ to be zero as is the case in STEP1. 
In this case there will be 
$m_i$ and $c_i$ values associated with $\gamma_1$ and $\gamma_2$; $m_1$, $m_2$, $c_1$, $c_2$.

To implement the {\sc lensfit} method we used {\sc SExtractor} (Bertin \& Arnouts, 1996),
on each set of PSF images to create 
a catalogue for the rotated and unrotated sets of images, we then 
create a matched catalogue in which only galaxies that were detected in both rotated and unrotated 
catalogues are kept. For each PSF the 
positions of the galaxies are the same over every shear value. We measured the PSF from the 
starfield images by using {\sc SExtractor} to identify the star positions. For PSFs D and E we 
also used the stars which were detected in the galaxy images and co-added this to the PSF from the 
starfield since a poor characterisation of these highly elliptical PSFs could affect the shear 
found from these sets of images as seen in Massey et al. (2007).

For the global shear estimates we used every galaxy in the matched catalogues to determine the 
intrinsic ellipticity prior from the zero-shear image provided for each PSF i.e. the prior 
was averaged over all size and magnitude ranges. For the investigation into the size and magnitude 
dependence of the estimated shear using these simulations we re-created the prior for each size and
magnitude bin using only the galaxies in that bin. We found that the prior exhibited significant 
variation over the magnitude and size ranges investigated. 

We calculate the errors on $\gamma^M_i-\gamma^T_i$ and hence the best fit values of  $m_i$ and $c_i$ 
with associated errors in the same way as 
for STEP1, described in Section \ref{Application to the STEP1 Simulations}. 
This results in a most likely value for $m_i$ and $c_i$ 
for each PSF with associated errors, Figure \ref{gammas} shows the linear
fit to $\gamma^M_i-\gamma^T_i$ for the PSF A set of shear values.
The average $\langle m\rangle$ and $\langle c\rangle$ 
is taken over all the values from each PSF and over $\gamma_1$ and 
$\gamma_2$. The error presented 
on the average is the same as presented in Massey et al. (2007) which is the 
\emph{average of the errors} over all PSFs
\be
\label{averror}
\sigma(\langle m\rangle)=\frac{\sum_{\rm psf}\sigma(m_{\rm psf})}{N_{\rm psf}}
\ee
where $N_{\rm psf}$ is the number of PSFs. 
This is meant to produce an error which is indicative of the expected error that one should get  
when using a particular shape measurement method on a given data set.  
\\

\noindent{\bf Bias and Offset for the Whole STEP2 catalogue}
\\

For the analysis of the entire catalogue we make no additional size or magnitude cuts other 
than those implicit in the {\sc SExtractor} source extraction, we use 
every galaxy in the matched catalogue for each image in each PSF set. 
\begin{figure}
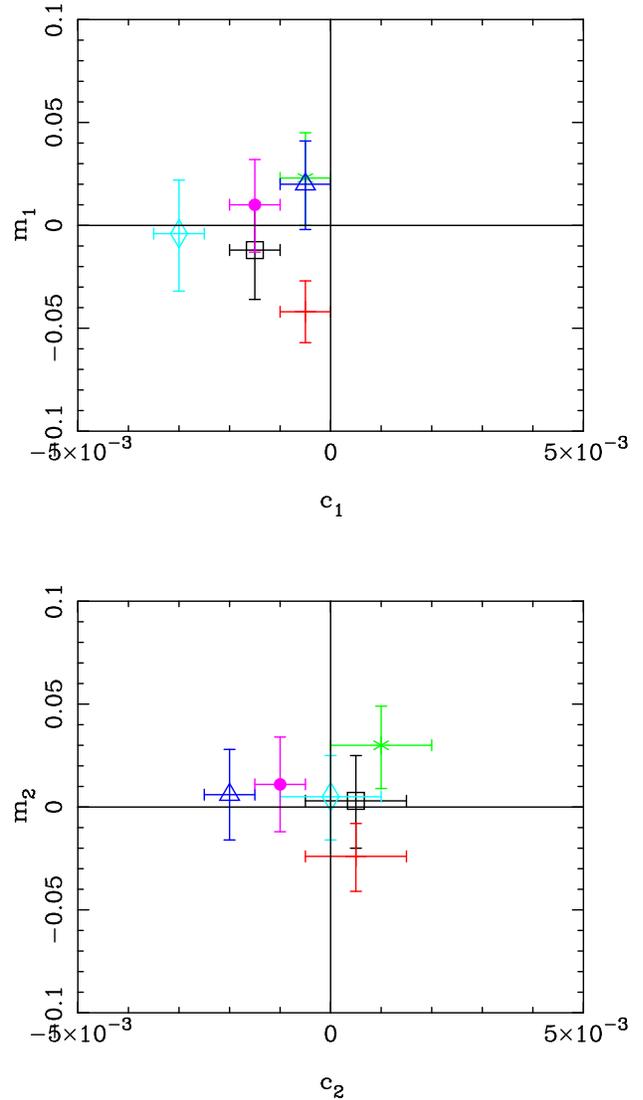

  \resizebox{84mm}{!}{
    \includegraphics{step2_scatter_1.ps}}

  \vspace{2mm}

  \resizebox{84mm}{!}{
    \includegraphics{step2_scatter_2.ps}}

 \caption{The top panel shows the best fit $m_1$ and and $c_1$ values with errors 
   for the STEP2 simulations. The bottom panel shows the best fit $m_2$ and and $c_2$ values.
   In both panels Black ($\Box$)=PSF A, Red ($+$)=PSF B, Green ($\times$)=PSF C, 
   Blue ($\bigtriangleup$)=PSF D, Cyan ($\Diamond$)=PSF E, Magenta ($\bullet$)=PSF F.
 }
 \label{step2_scatter}
\end{figure}
Figure \ref{step2_scatter} shows the best fit $m_i$ and $c_i$ values for $\gamma_1$ and $\gamma_2$ 
for each PSF. It can be seen that there is no general pattern or offset in the values or 
any correspondence between the $\gamma_1$ and $\gamma_2$ values for any particular PSF. This is 
as expected since the points should be randomly scattered about ($m=0$, $c=0$) with a 
dispersion due to the finite size of the galaxy sample. The value of $c_1$ is slightly 
systematically offset from $c_1=0$, we will discuss this further later in this Section. 
Figure \ref{step2_final} shows 
that the scatter in bias is indeed statistical since when averaging over all PSFs the value of 
$\langle m\rangle\sim 0.002$, the results are presented in detail 
in Table \ref{STEP2table}. This shows that the {\sc lensfit} method has a smaller bias 
of any method presented in the STEP2 publication (Massey et al., 2007). 
Furthermore, the most likely values of 
$m$ and $c$ do not vary substantially when PSF D and E, which have 
the strongest PSF distortions, are removed. This suggests that the scatter 
in Figure \ref{step2_scatter} is indeed purely statistical. Note that the error bars do not increase 
since they are the average errors on $m$ and $c$ for the PSFs used, see equation (\ref{averror}).
\begin{figure}
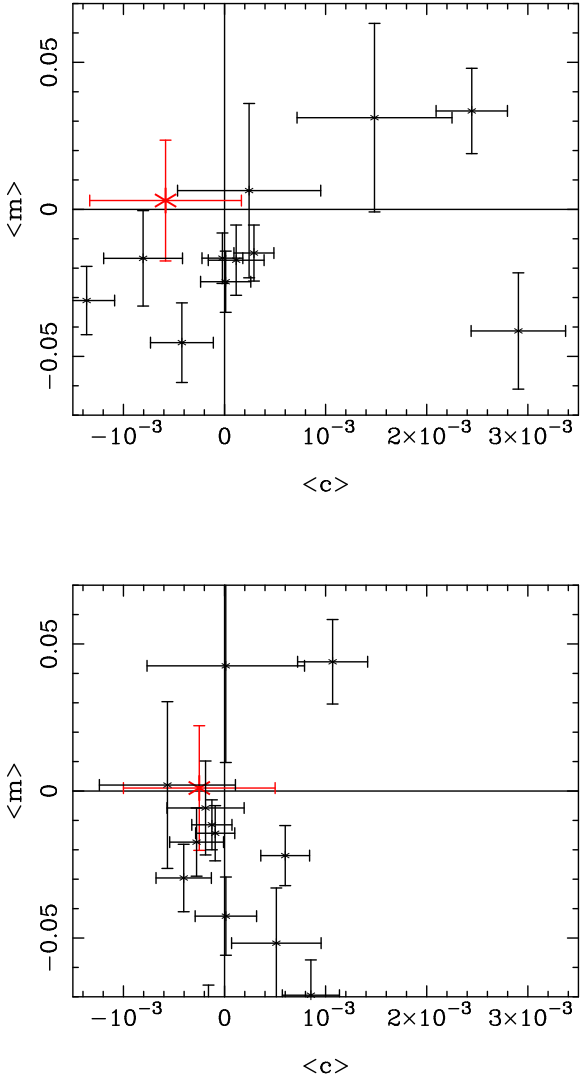

\resizebox{84mm}{!}{
\includegraphics{step2_allpsf_final.ps}}

\vspace{2mm}

\resizebox{84mm}{!}{
\includegraphics{step2_noellippsf_final.ps}}
\caption{Adapted from Massey et al. (2007), Figure 6. The red
  points (in larger font) show the result of using {\sc lensfit} on the STEP2 simulations. The black 
  points show other shape measurement methods analysed in Massey et al. (2007). The top 
  panel shows the value of $m$ and $c$, averaged over all PSFs and 
  $\gamma_1$ and $\gamma_2$. 
  The bottom panel shows the value of $m$ and $c$ averaged 
  over PSFs A, B, C and F i.e. with the highly elliptical PSFs D and E ignored. The errors 
  shown are the average of the errors on each PSF, equation (\ref{averror}).}
\label{step2_final}
\end{figure}

The slightly larger value of 
$\langle c\rangle$ relative to the other STEP2 methods 
is most likely due to residuals in the PSF estimation. We make this 
assertion since a systematic error on 
PSF estimation is the most straightforward way to create a non-zero $c$ value and 
also because we have identified pixelisation of the PSF as a potential source of limitation.  
The non-zero $c$ value 
is not a large concern for two reasons. Firstly it has been shown (for example in 
Amara \& Refregier, 2007; Kitching et al., 2008a) that it is the bias $m$ not an offset $c$ 
in the estimated shear that has the largest effect on cosmological parameter estimation. 
Secondly the way in 
which the PSF is determined is not central to the method, for example any PSF determination 
routine could be used 
in conjunction with the unbiased shear estimation method to reduce the 
$c$ value. Furthermore planned space-based 
wide field imagers such as DUNE and SNAP will have very stable PSF 
modelling at high resolution before launch. 
\begin{table*}
\begin{center}
\begin{tabular}{|l|c|c|c|c|c|}
\hline
Data Set&Galaxy Sample&$\langle m\rangle$&$\sigma(\langle m\rangle)$&
$\langle c\rangle$&$\sigma(\langle c\rangle)$\\
\hline
{\bf STEP1}&&&&&\\
\hline
All PSFs&SExtractor catalogue&$+0.0058$&$0.0056$&$-0.0006$&$0.0002$\\
\hline
{\bf STEP2}&&&&&\\
\hline
All PSFs&SExtractor matched catalogue&$+0.0020$&$0.0205$&$-0.00071$&$0.00066$\\
No PSF D \& E&SExtractor matched catalogue&$+0.0010$&$0.0211$&$-0.00025$&$0.00075$\\
\hline
All PSFs&$18\leq$ Mag $< 20$&$-0.0640$&$0.1099$&$-0.0029$&$0.0028$\\
All PSFs&$20\leq$ Mag $< 21$&$-0.0167$&$0.1120$&$+0.0020$&$0.0029$\\
All PSFs&$21\leq$ Mag $< 22$&$+0.0134$&$0.0155$&$+0.0015$&$0.0005$\\
All PSFs&$22\leq$ Mag $< 23$&$+0.0019$&$0.0158$&$-0.0011$&$0.0005$\\
All PSFs&$23\leq$ Mag $< 24$&$-0.0177$&$0.0158$&$-0.0003$&$0.0005$\\
All PSFs&$24\leq$ Mag $< 25$&$-0.0049$&$0.0231$&$-0.0033$&$0.0006$\\
\hline
All PSFs&$0.4\leq$ Radius $< 0.6$&$+0.0094$&$0.0061$&$-0.0015$&$0.0003$\\
All PSFs&$0.6\leq$ Radius $< 0.8$&$+0.0031$&$0.0169$&$+0.0001$&$0.0005$\\
All PSFs&$0.8\leq$ Radius $< 1.0$&$-0.0192$&$0.0232$&$-0.0005$&$0.0007$\\
All PSFs&$1.0\leq$ Radius $< 1.2$&$-0.0130$&$0.0315$&$-0.0041$&$0.0008$\\
\hline
\end{tabular}
\caption{The STEP1 and STEP2 $m$ and $c$ results. We use galaxies in the catalogues created using 
{\sc SExtractor}, for STEP2 we match the rotated and unrotated catalogues. 
The average number density of galaxies over all PSFs used in the STEP1 analysis was $9$ 
per square arcminute, for STEP2 the matched number density was $30$ per square arcminute.}
\label{STEP2table}
\end{center}
\end{table*}
\\

\noindent{\bf Bias and Offset as a Function of Size and Magnitude}
\\

\begin{figure}
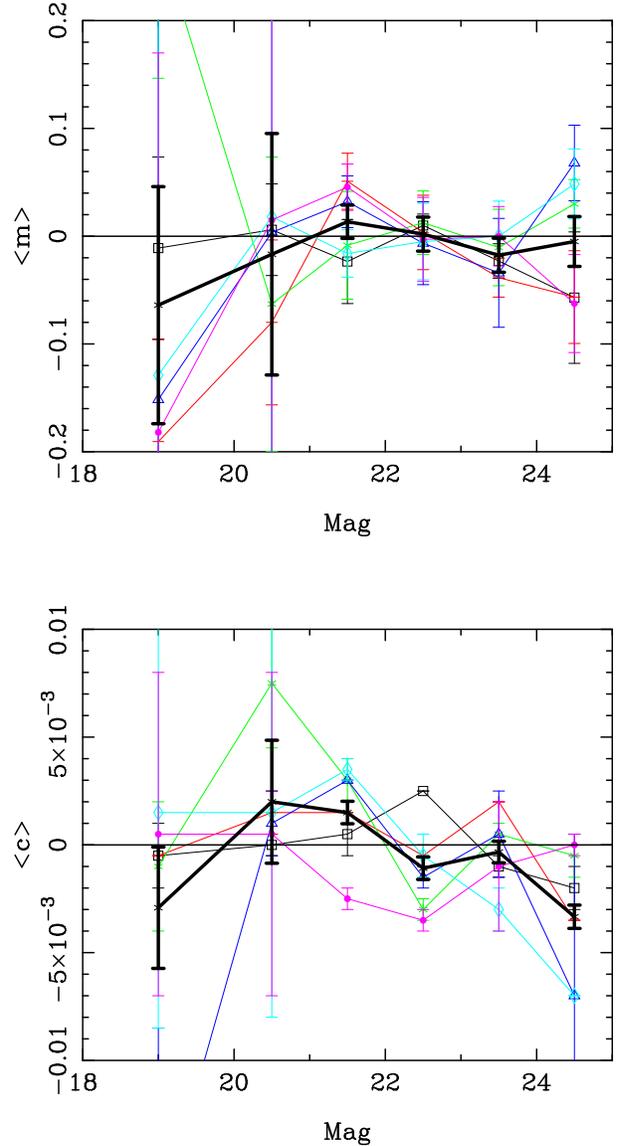

  \resizebox{84mm}{!}{
    \includegraphics{step2_mag_m.ps}}

  \vspace{2mm}

  \resizebox{84mm}{!}{
    \includegraphics{step2_mag_c.ps}}

 \caption{The variation in the bias $\langle m\rangle$ and offset $\langle c\rangle$ 
   as a function of magnitude for the STEP2 analysis. The fainter coloured lines are the values 
   from each individual PSF; 
   in both panels for the fainter lines 
   Black ($\Box$)=PSF A, Red ($+$)=PSF B, Green ($\times$)=PSF C, 
   Blue ($\bigtriangleup$)=PSF D, Cyan ($\Diamond$)=PSF E, Magenta ($\bullet$)=PSF F. The 
   bold black lines show the average over all PSFs. The points in magnitude are at the centre of the 
   bin used, see Table \ref{STEP2table} for the values of the bin boundaries used. 
 }
 \label{step2_magnitude}
\end{figure}

Here we show how the bias and offset vary as a function of 
magnitude and size, the detailed results are summarised in Table \ref{STEP2table}. 
The {\sc SExtractor} matched 
catalogues used were set to the zero-point of $M=30.8$ as discussed in Massey et al. (2007). 
The value of the radius of the galaxy uses a shapelet based definition (Massey \& 
Refregier, 2005; equation 53) these values were provided for each galaxy we detected from the 
STEP2 website\footnote{http://www.physics.ubc.ca/$\sim$heymans/step/step2\_info.html}; note that we 
did not get positions from the website but only radii for galaxies that we had detected using 
{\sc SExtractor}. 
We re-iterate that the intrinsic ellipticity prior was recalculated 
for each magnitude and size bin, always assuming a zero-sheared functional form as described 
in Section \ref{Estimation of the Prior}. Figure \ref{step2_magnitude} shows how the bias 
$m$ and offset $c$ vary as a function of the magnitude of the galaxies used. The fainter coloured 
lines show how this varies for each individual PSF, averaged over $\gamma_1$ and $\gamma_2$, 
the bold black lines show the average 
$\langle m\rangle$ and $\langle c\rangle$ over all PSFs. 

Similarly to the average of $m$ and $c$ 
when using the whole sample there is a scatter of values from each PSF about the 
$\langle m\rangle=0$ line, however this is dominantly statistical since when taking the average the 
value of the bias is $|\langle m\rangle|< 0.02$ for $20<M<24$. The deviation at $M<20$ is due to 
the number of galaxies in this bin being small however the error 
bars show that none of the points' variation from $\langle m\rangle=0$ is statistically 
significant. 
The weak variation of $\langle c\rangle$ as a function of magnitude shows that the method is 
robust to the magnitude range used. 
The only statistically significant variations occur where the number of galaxies in the bin 
become very low i.e. $M<20$ and $M>24$. The method performs better in certain magnitude 
bins than when the 
sample is taken as a whole, this is because the intrinsic ellipticity prior which is recalculated 
for each magnitude bin now better represents the intrinsic distribution of ellipticities in that bin. 
The variation of the intrinsic ellipticity distribution was an issue highlighted in the STEP2 
publication Massey et al. (2007).  
By taking a global average this information is averaged over so that the global prior is less 
representative of some galaxy sub-populations. This highlights the need in an actual implementation 
of this method to calculate the prior as a function of galaxy property; in STEP2 this was only 
magnitude and size but this could be extended to colour and galaxy-type. 

The small number of galaxies in the ranges $M<20$ and $M>23$ has a dominant effect
on the determination of the prior as discussed in Section \ref{Estimation of the Prior}. It is 
encouraging that despite a poorly estimated prior in these bins the variation is still small 
$|\langle c\rangle|\ls 0.003$ and that the bias $\langle m\rangle$ is unaffected. 
In the STEP2 simulation we are using a total sample of $\sim 1500$ galaxies with 
which to estimate the prior, and when splitting this sample into size and magnitude this number 
becomes much less especially at the extremes of the magnitude and size values. In real weak lensing 
surveys the number of galaxies is at least $\gg 10^4$ so that any cut in size or magnitude should 
contain $\gg 100$ galaxies which is enough to accurately determine the prior. 
 
\begin{figure}
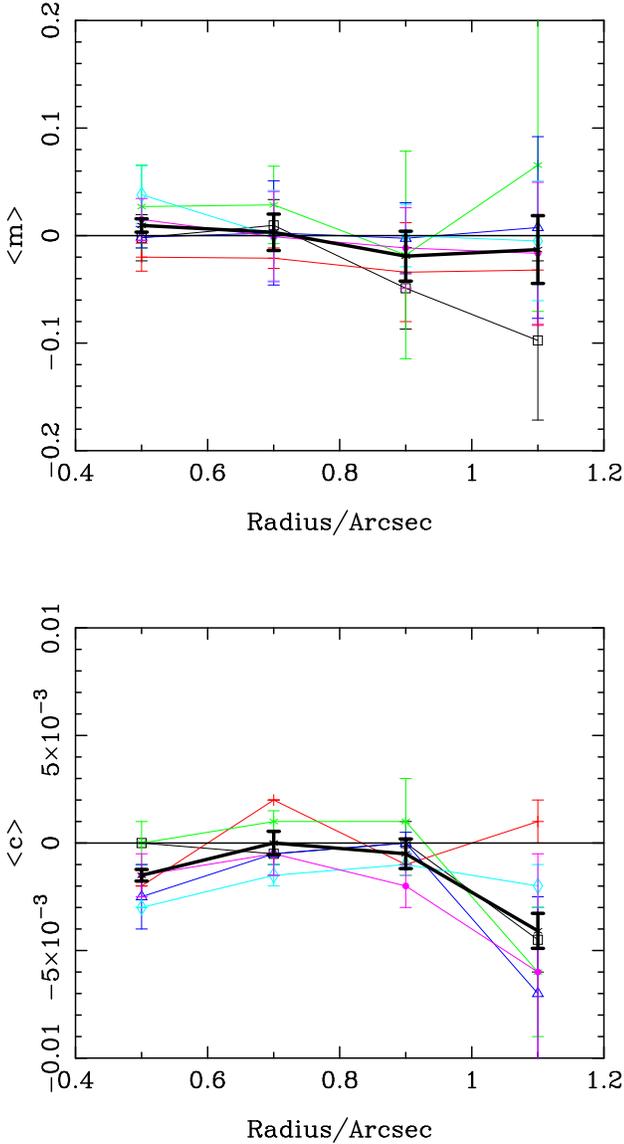

  \resizebox{84mm}{!}{
    \includegraphics{step2_rad_m.ps}}

  \vspace{2mm}

  \resizebox{84mm}{!}{
    \includegraphics{step2_rad_c.ps}}

 \caption{The variation in the bias $\langle m\rangle$ and offset $\langle c\rangle$ 
   as a function of radius for the STEP2 analysis. The radius is shown in units of arcseconds, where 
   the pixel scale of the STEP2 simulations is $0.2$ arcseconds per pixel. 
   The fainter coloured lines are the values 
   from each individual PSF; 
   in both panels for the fainter lines 
   Black ($\Box$)=PSF A, Red ($+$)=PSF B, Green ($\times$)=PSF C, 
   Blue ($\bigtriangleup$)=PSF D, Cyan ($\Diamond$)=PSF E, Magenta ($\bullet$)=PSF F. The 
   bold black lines show the average over all PSFs. The points in radius are at the centre of the 
   bin used, see Table \ref{STEP2table} for the values of the bin boundaries used.
 }
 \label{step2_radius}
\end{figure}

Figure \ref{step2_radius} shows the variation in $\langle m\rangle$ and $\langle c\rangle$ 
as function of galaxy radius, as in  
Figure \ref{step2_magnitude} the fainter lines show the values for each PSF individually and the 
bold line shows the average over all PSFs. The variation in the bias over the whole range in 
radius is $|\langle m\rangle|\ls 0.02$ with no point being a statistically significant deviation from 
$\langle m\rangle=0$.  
The variation of the offset as a function of magnitude is very small in the 
range $0.6<$Radius$<1.0$ arcseconds. Again, the method performs better in certain radius bins than 
when the galaxy sample is taken as a whole, we re-iterate that this is due to the prior 
being a better representation of the intrinsic distribution of ellipticities in each bin. 

The deviation at $<0.6$ arcseconds and $>1.0$ arcseconds is again due to the 
small number of galaxies in each bin per PSF, this affects the 
determination of the ellipticity prior as discussed previously.  Again even with this small 
number of galaxies the bias is unaffected.
Note that the STEP2 pixel scale is $0.2$ arcseconds/pixel so that galaxies in 
the bin $0.4<$Radius$<0.6$ arcseconds span only $\sim 2$ -- $3$ pixels.  

\subsection{Discussion}
\label{Discussion}
In the previous Section \ref{Application to the STEP1 Simulations} we presented the results of using 
the {\sc lensfit} method on the STEP1 and STEP2 simulations. The performance of the method is 
parameterised by calculating the difference between the input shear value for a given image and the 
estimated shear from that image. This quantity is then fitted, as a function of input shear, with 
a linear function. The function is parametrised by a bias $m$ and an offset $c$, defined in equation 
(\ref{M_c}). The results of this application to the simulations are summarised in detail in Table 
\ref{STEP2table}.

We found that the method performed very well in comparison to the other methods 
presented in the STEP publications. In particular the bias $m$ is smaller in both the STEP1 and STEP2 
simulation results than the vast majority of methods, and performs consistently well over the whole 
suite of simulations. The small residual bias in the STEP1 simulation could be attributed to 
inaccuracies in the PSF characterisation due to pixelisation effects, 
see Section \ref{PSF Estimation}. 
In STEP2 we found again that the bias $m$ was small in comparison to other methods. Furthermore when 
two of the PSFs were removed, the most elliptical PSFs (D and E) the best fit values of the $m$ 
and $c$ values change by a very small margin, this suggested that the scatter in the values, shown in 
Figure \ref{step2_scatter} 
is entirely statistical and due to the finite, and small, number of galaxies in each STEP2 image. 
It has been shown in Kitching et al. (2008a) and Amara \& Refregier (2007) that when using 
cosmic shear it is a bias, not
an offset, in the shear estimation that has the largest effect on cosmological parameter estimation. 
The small offset that we find in $c$ is not significant since the bias of the method is small. 
Furthermore when we correctly recalculate the bias as a function of magnitude and size the 
small offset remains consistent with $c\approx 0$. 

We now refer to Table 1 in Massey et al. (2007). 
To summarise, all PSFs use shapelet galaxies except PSF 
B which uses exponential profiles, PSF A and B are the same SUBARU 
PSF but use different galaxy types. PSF 
C is an enlarged PSF, and PSFs D and E are highly elliptical aligned along the $x$ ($e_1$) and 
$45^{\circ}$ ($e_2$) axes respectively. 
PSF F is circularly symmetric. It can be seen from Figure \ref{step2_scatter} 
that there is no pattern in the best fit values of the bias $m$ and offset $c$ as function of galaxy 
type or PSF. The PSF for which these values is largest is PSF C, this is most likely due to a 
sub-optimal characterisation of the PSF. There is no significant difference between
the exponential or shapelet simulated galaxy sets. Even though we have assumed a de Vaucouleurs 
galaxy profile the method retains its ability to fit this model to either shapelet, exponential or 
bulge plus disk (STEP1 i.e. exponential disk plus a de Vaucouleurs bulge). This is because 
differences in the surface brightness profiles are subtle and not significant at low signal-to-noise. 
Also the exact radial profile should not matter too much at faint magnitudes
since it is effectively unconstrained.
The method could be extended to fit to individual nodes of sub-structure in galaxies 
with complex morphologies, and the exact form of the model profile used is not a central 
tenant of {\sc lensfit} method, however since the vast majority of galaxies used in cosmic shear 
analysis will be faint we expect that either a de Vaucouleurs or 
an exponential profile will suffice. 

By 
calculating the bias and offset as a function of size and magnitude, Figures \ref{step2_magnitude} 
and \ref{step2_radius}, 
we have shown that the bias $m$ 
remains at $|m|<0.02$ 
over a wide range in size and magnitude. Furthermore the offset $c$ is consistent 
with $0$ in the regime that there are a sufficient number of galaxies to estimate the intrinsic 
ellipticity prior, see Section \ref{Estimation of the Prior}. The only statistically significant 
deviation in the offset occurs where the number of galaxies available in the analysis becomes 
small which makes the estimation of the prior difficult since the the ($e_1$, $e_2$) 
plane is not sampled to sufficient accuracy. We stress that in the case of real data this will not 
be a problem, as discussed in Section \ref{Application to the STEP2 Simulations}.

We emphasise here that although this analysis has been carried out after the details 
of the STEP simulations were made public we did not iterate on the STEP1 or STEP2 
simulations to tune any \emph{ad hoc} parameters or vary the shape measurement method. The numerical 
convergence of the parameter space values were found using the zero-shear image from STEP1 PSF0. 
In our investigation we did however 
find some nuances of the STEP simulations 
which we will highlight. For STEP1 we found that the intrinsic ellipticity prior is very sharply 
peaked about zero and that the functional form used in the prior needs to be 
sufficiently able to fit this peak. 

In STEP2 we found that when analysing the PSFs D and E we found that the shear values 
obtained from these 
images were large compared to the expected shear value (up to $5\%$ bias, and up to 
$0.01$ shear offset). We found that to 
fully characterise the PSF for these sets required more stars than just the ones 
in the starfields to determine the PSF, to yield an accurate PSF we 
co-added the PSF derived from the starfield and galaxy fields. Furthermore the PSF 
determined from the starfield, rotated image 
and unrotated image separately were not fully compatible i.e. they 
varied to such a degree that the shear bias and offset could be affected by up 
to $5\%$ if either one of the PSF's (from the starfield or galaxy images) 
were used individually. To resolve this issue we did iterate on this data, but only 
to discover problems in the input data not the shape measurement method itself. 
Note that the small bias and offset presented
are not dependent on this aspect of the data since when PSF's D and E are removed the 
{\sc lensfit} method still finds a smaller bias and offset than any other method in the STEP2 
publication (Massey et al., 2007). 
We also found that the starfields for PSFs A, B and C contained 
no noise. 

\section{An alternative to the STEP parametrisation}
\label{Beyond m and c values}
We will now investigate the results going beyond the $m$ and $c$ parameterisation. These 
results do not only compare the absolute values of some quantity relative to the `ideal' result, 
of $m=0$ and $c=0$ for example, but will assess whether any deviation in the estimated shear values 
found by applying a shape measurement  method to simulations is statistical, due to the finite 
number of galaxies, or is a property of the method. 

The statistic will also use more information than the STEP parameterisation. As can be seen from Figure \ref{gammas} the $m$ and $c$ 
parameterisation is well suited to STEP1 in which the number of points tested in shear is small, and a linear parameterisation 
can capture most of the relevent information. However, when using fewer galaxies per shear value (so that the variance is larger) 
and using many more shear values as in STEP2, the $m$ and $c$ parameterisation disregards a large majority of the information by fitting 
a simple linear function through many noisy points. The approach presented here is well suited to STEP2-like simulations in which there 
are many shear values for which a relative large variance is expected. 

\subsection{The Quality Factor}
Bridle et al. (2008) (GRavitational lEnsing Accuracy Testing, GREAT08 Handbook) 
define a Quality factor Q which 
allows one to compare the expected statistical distribution of estimated shear values from 
a simulation with the 
distribution measured by a method. The GREAT08 Quality factor is based on the work done in 
Amara \& Refregier (2007) on the desired minimum statistical and systematic 
spread of estimated shear values when designing 
a future weak lensing survey. In this paper we present a generalisation of the GREAT08 
Quality factor for use in an arbitrary weak lensing simulation. The central variable used here 
is the same as that used in the $m$ and $c$ 
analysis, which is the difference between the estimated shear and the input shear 
$\gamma^M_i-\gamma^T_i$. For a good shear measurement method, that contains no biases, 
the variance in this quantity should be entirely statistical. The quantity used is the 
average mean-square error $\langle (\gamma^M_i-\gamma^T_i)^2\rangle$. 
The statistical spread from the  
simulation in question is denoted by $\sigma^2_{stat}$. This expected variance is 
related to the measured spread of values via the `Quality factor' $Q$ which we define as 
\be
Q=1000\frac{\sigma^2_{stat}}{\frac{1}{2}\frac{1}{N_{\rm images}}\sum_{i=1,2}\sum_{\rm images}\langle (\gamma^M_i-\gamma^T_i)^2\rangle}
\ee
where the mean-square error is averaged over $\gamma_1$ and $\gamma_2$ for each image (input shear 
value) in a simulation. The factor of $1000$ normalises the expression so that a method which 
performs well should have $Q\sim 1000$ i.e. the spread in estimated shear is purely statistical.
Note that the nominator $\sigma_{stat}$ is the shear variance of the galaxies analysed and is set by 
the simulations. 
The Quality factor averages over all values of $\gamma^T$ in an analogous way to the $m$ and $c$ 
parameterisation, which fits a functional form to $\gamma^M_i-\gamma^T_i$ over all values of 
$\gamma^T$. This effectively averages over the angular scale on which shear is averaged as we shall discuss.
 
The mean square error can be written as 
a sum of the intrinsic variance and a bias $\langle (\gamma^M-\gamma^T)^2\rangle=
\langle(\gamma^T)^2\rangle+[{\rm Bias}(\gamma^T,\gamma^M)]^2$ where 
${\rm Bias}(\gamma^T,\gamma^M)=\langle\gamma^M\rangle-\gamma^T$ so that the Quality factor 
effectively parameterises any residual bias in the estimators $\gamma^M$; for an unbiased estimator 
the mean-square-error is equal to the variance of the data. This is an example of a 
\emph{loss function} that parameterises the amount that an estimator differs 
from an underlying distribution. The mean-square error penalises outliers due to the quadratic nature of the
function, an example of a loss function that does not penalise outliers to such a degree is the 
absolute loss function $\langle|\gamma^M-\gamma^T|\rangle$. This  
loss function could also be 
used to make effective comparisons between the shear estimations from several different shape measurement 
methods, for a good shear estimator the absolute loss function should be close to zero. 

When \emph{designing} a simulation and considering what value of the Quality factor would render 
a shape measurement method 
`adequate' (for use in current or future surveys) one must define the variance in shear that a 
particular survey requires, 
$\sigma$. In Bridle et al. (2008) the nominator in the GREAT08 Quality factor is effectively 
$10^{-4}=1000\times \sigma^2$ where $\sigma^2=\sigma^2_{stat}+\sigma^2_{systematic}$ 
is the sum of expected statistical and systematic errors, and 
so slightly differs from the definition presented in this paper.
When designing a simulation the requirement of a particular shear 
variance defines the number of galaxies $N$ in the 
simulation via $\sigma_{stat}=\sigma_{\epsilon}/\sqrt{N}$. This is justifiable since one 
can determine the shear variance 
that a particular survey will need in order to 
fully utilise the data (van Waerbeke et al., 2006; Amara \& Refregier, 2007)
and create a simulation that allows one to simulate the expected data. 

\subsection{Relation to the STEP parameterisation}
The relationship between the Quality factor and the STEP parameterisation is not straightforward and one 
should exercise caution when making a mapping between the two statistics, as we shall discuss. 
A subtlety also arises in the scale dependence of the statistics when one considers the level of bias 
or offset that one requires for a future survey and attempts 
to determine the requirement on the Quality factor that this would imply. 

The Quality factor effectively combines, in a non-trivial but justifiable way, the information 
from the four STEP parameters: $m$, $c$ and the uncertainties on these values $\Delta m$ and $\Delta c$. 
As such one must take care when determining a Quality factor from the absolute $m$ and $c$ values alone,  
in fitting this linear functional form, information on a method's performance is 
lost due to the assumption of the functional form itself. 
By using equation (\ref{M_c}) one can relate the 
Quality factor to the STEP $m$ and $c$ parameterisation (for clarity in the following we let 
the angular brackets correspond to 
the averaging over shear, images as well as $\gamma_1$ and $\gamma_2$) 
the average values are calculated by integrating the function in the angular brackets over the 
interval $-\gamma_L$ to $+\gamma_L$ where $\gamma_L=0.06$ in the STEP2 simulations,
\ba
\label{STEPQ} 
Q&=&1000\frac{\sigma^2_{stat}}{\langle m^2(\gamma^T)^2 + 2cm\gamma^T + c^2\rangle}\nn
&=&1000\frac{3\sigma^2_{stat}}{m^2\gamma^2_L+3c^2},
\ea
where we have assumed that the true shear values are evenly distributed in the range $-\gamma_L$ to 
$+\gamma_L$ i.e. $P(\gamma)=$constant for all $|\gamma|\leq\gamma_L$. 
In the case that $c=0$ the Quality factor is simply inversely proportional to $m^2$.
This highlights the difference between the Quality factor and the STEP parameterisation,  
given a simulation the STEP parameterisation quantifies a methods performance by the 
$m$ and $c$ values with the hope that $m\sim 0$, but this does not quantify whether 
such values achieved are statistically significant, the Quality factor 
essentially combines the bias and offset along with the uncertainties on these values into a single 
parameter. 
As an aside we note that the substitution of the STEP parameterisation into the absolute loss function 
gives $\langle|\gamma^M-\gamma^T|\rangle=|c|$. 

As shown (most recently by Fu et al. 2007; Figure 5), $\langle(\gamma^T)^2\rangle$ varies as a 
function of angular scale. 
So by choosing an average value of $\langle(\gamma^T)^2\rangle$ one implicitly 
assumes that $m$ is averaged over scale.
If a particular value of $\langle(\gamma^T)^2\rangle$ is chosen (as opposed to taking the average) 
then this corresponds to picking a certain scale over which shear variance is averaged. 
Furthermore a degeneracy exists when 
determining the required Quality factor between $m$, $c$ and scale. This can 
be seen by referring to Fu et al. 2007 (Figure 5): if $c > 0$ then for a particular value of 
$\langle(\gamma^T)^2\rangle$ the scale to which this corresponds to will increase. 
The exact 
relation between the bias, offset and scale will depend on the simulation 
through $\sigma_{stat}$. 
This bias, offset and scale degeneracy highlights 
the fact that the Quality factor 
itself averages over scale, but that this is no more pernicious than the STEP 
parameterisation in this regard. 

We emphasise that using the STEP $m$ and $c$ 
values to calculate a Quality factor using 
equation (\ref{STEPQ}) merely gives the maximum possible Q for those $m$ and $c$ values. 
\begin{figure}
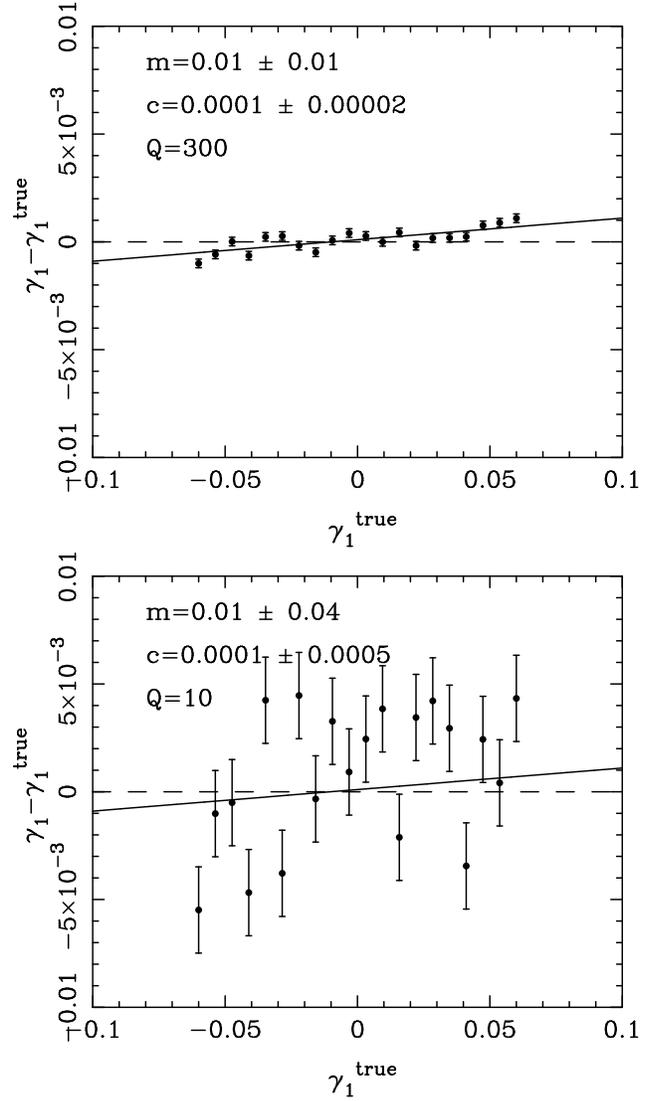

  \resizebox{84mm}{!}{
    \includegraphics{m_c_300.ps}}

  \resizebox{84mm}{!}{
    \includegraphics{m_c_3.ps}}

 \caption{Simulated results of measured shear showing $\gamma^M-\gamma^T$ for two different 
   realisations. Both panels show results which have best fit $m$ and $c$ values of $m=0.01$ and $c=0.0001$ 
   the solid lines show this fit. However the Quality factor of the two results is very different 
   due to the scatter of points about the best fit line. The maximum Quality factor for these $m$ and $c$ 
   values would occur if all the points lay exactly on the best fit line, 
   using equation (\ref{STEPQ}) is $Q=770$. Note that in the two cases the uncertainty on the 
   best fit $m$ and $c$ values are different, the Quality factor effectively combines the best fit values 
   and the uncertainties into a single parameter. 
 }
 \label{STEPQfig}
\end{figure}
The two panels in Figure \ref{STEPQfig} show that for the same $m$ and $c$ values the Quality factor can be 
very different (for these we assume that $\sigma^2_{stat}=10^{-7}$). 
Using equation (\ref{STEPQ}) the Quality factor found using these values would be 
$Q=770$, however this would only be 
achieved if all the points in Figure \ref{STEPQfig} had zero scatter about the best fit line. The Quality 
factor thus takes into account both the bias and offset as well as the scatter of points. 
However as can be seen from equation (\ref{STEPQ}) different sets of $m$ and $c$ values can 
produce the same Quality factor. 

In the STEP2 and GREAT08 simulations $|\gamma^T|\leq 0.06$ so that 
$\langle(\gamma^T)^2\rangle=(1/3)(0.06)^2\sim (0.03)^2$.   
If we assume that $\sigma^2_{stat}\sim 10^{-7}$ and $m\sim 0.1$, as is found in STEP2 
when investigating magnitude and size dependence of the methods, it can be seen that existing 
methods have a Quality factor of $Q\ls 10$ which is sufficient for current surveys 
(see Heymans et al., 2006; and the hatched region in Figure \ref{step1_final}). 
As discussed in Bridle et al. (2008) 
if a method only recovers a single constant value of zero shear for any input shear value, 
$\gamma_1 = \gamma_2 = 0$ then $Q\sim ~0.1$. We re-iterate that comparing $m$ and $c$ values 
with the Quality factor a limit 
is inevitably reached since in fitting the STEP parameterisation 
to a large number of points as in STEP2 
information on the scatter of the points is lost in the fitting process.  

We stress that the issues with the Quality factor that were previously discussed  
will only arise when designing a simulation and assessing which Quality factor corresponds to a 
particular bias or offset 
requirement. When presented with existing simulations one can readily calculate the Quality 
factor which allows 
the shear variance of a method to be compared to the intrinsic shear variance of the simulation. 

\subsection{Determination of the Quality Factor from the STEP2 Simulation}
\begin{figure}
  \resizebox{84mm}{!}{
    \includegraphics{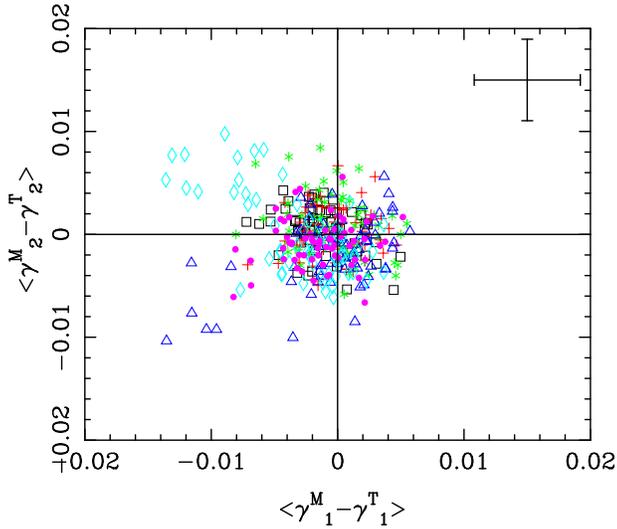}}
  \caption{The values of $\gamma^M_1-\gamma^T_1$ and $\gamma^M_2-\gamma^T_2$ for all PSFs from the 
    STEP2 simulations. The Figure shows a scatter in the values about zero, the scatter is due to 
    intrinsic variance due to the number of galaxies in the simulation, the scatter about zero 
    shows that the shear offset is small. The error bars show the average error on each point. The 
    colours and symbols again represent the various PSFs with 
    Black ($\Box$)=PSF A, Red ($+$)=PSF B, Green ($\times$)=PSF C, 
    Blue ($\bigtriangleup$)=PSF D, Cyan ($\Diamond$)=PSF E, Magenta ($\bullet$)=PSF F.
 }
 \label{step2_Q}
\end{figure}

Figure \ref{step2_Q} shows the spread in $\gamma^M_i-\gamma^T_i$ for $\gamma_1$ and $\gamma_2$ 
for the {\sc lensfit} application to the STEP2 simulations. We will not show results for $Q$ from 
STEP1 since the number of points is so small (only $5$ shear values) 
that results on $Q$ may be inaccurate.
In Figure \ref{step2_Q} $c\not=0$ would mean that the points would be scattered 
about a point offset from the origin, $m\not=0$ would mean the spread of the points about zero 
would be larger than the intrinsic shear variance of the STEP2 simulation.  

It can be seen from Figure \ref{step2_Q} that there is a spread in estimated shear values about 
($\gamma^M_1-\gamma^T_1\approx 0$, $\gamma^M_2-\gamma^T_2\approx 0$), that is expected for a 
method which can 
accurately estimate the shear. The points which are scattered furthest from the origin 
are all associated with the highly elliptical PSFs D and E. 
Usually the expected statistical mean-square error would be given, assuming Poisson statistics, by 
$\sigma_{stat}^2=\langle (e_i^{\rm int})^2\rangle/N$. However 
as discussed in Section \ref{Application to the STEP2 Simulations} the statistical 
error for the STEP2 
simulations is reduced due to the co-addition of rotated and unrotated images to 
$\sigma^2_{stat}=\langle (e_i^{\rm int})^4\rangle/2N$ for the STEP2 simulations 
$\sqrt{\langle (e_i^{\rm int})^4\rangle}\sim 0.05$ and $N\sim 3000$ so that 
$\sigma^2_{stat}\sim 4.2\times 10^{-7}$.

We find that for the {\sc lensfit} application to the STEP2 
simulations the global average value of 
$\langle (\gamma^M-\gamma^T)^2\rangle\sim 1.1\times 10^{-5}$ so that our global $Q$ factor is $Q=38$. 
This $Q$ factor shows that there is still some residual bias 
in the spread in the values of $\langle (\gamma^M_i-\gamma^T_i)^2\rangle$, we attribute this to 
poor estimation of the prior due to low numbers of galaxies at the extremes of magnitude and radius. 
If the highly elliptical PSFs are removed, PSFs D and E, then the $Q$ factor improves to $Q=58$. 
\begin{table*}
\begin{center}
\begin{tabular}{|l|c|c|c|}
\hline
Data Set&Galaxy Sample&$\sigma^2_{stat}$&$Q$\\
\hline
All PSFs&SExtractor matched catalogue&$4.2\times 10^{-7}$&$38.5$\\
No PSF D \& E&SExtractor matched catalogue&$4.2\times 10^{-7}$&$57.7$\\
\hline
All PSFs&$18\leq$ Mag $< 20$&$6.3\times 10^{-6}$&$45.3$\\
All PSFs&$20\leq$ Mag $< 21$&$3.2\times 10^{-6}$&$112$\\
All PSFs&$21\leq$ Mag $< 22$&$1.6\times 10^{-6}$&$93.8$\\
All PSFs&$22\leq$ Mag $< 23$&$1.6\times 10^{-6}$&$74.2$\\
All PSFs&$23\leq$ Mag $< 24$&$6.3\times 10^{-6}$&$295$\\
All PSFs&$24\leq$ Mag $< 25$&$3.1\times 10^{-5}$&$277$\\
\hline
All PSFs&$0.4\leq$ Radius $< 0.6$&$7.8\times 10^{-7}$&$33.3$\\
All PSFs&$0.6\leq$ Radius $< 0.8$&$1.6\times 10^{-6}$&$81.0$\\
All PSFs&$0.8\leq$ Radius $< 1.0$&$3.1\times 10^{-6}$&$89.0$\\
All PSFs&$1.0\leq$ Radius $< 1.2$&$1.3\times 10^{-5}$&$169$\\
\hline
\end{tabular}
\caption{The STEP2 Quality factor $Q$ for the global STEP2 analysis and a function of 
magnitude and size. STEP2 uses galaxies in the catalogues created using 
{\sc SExtractor} and matching the rotated and unrotated catalogues.}
\label{Qtable}
\end{center}
\end{table*}

We also show how the $Q$ value varies as a function of magnitude and size, the results are 
shown in Table \ref{Qtable}. When this is done the 
statistical variance $\sigma_{stat}$ is changed since there are 
fewer galaxies in the corresponding bins, this is 
shown in Table \ref{Qtable}. 

It can be 
seen for the variation in magnitude that the $Q$ values are generally higher, with an average 
$Q\sim 150$ than for the global 
sample, this is because the prior better represents the samples intrinsic ellipticity distribution 
in each bin. This is the same reason that the $m$ and $c$ values improve in some bins 
when the sample is split 
into size and magnitude bins, as discussed in Section \ref{Application to the STEP2 Simulations}. 
There is similar variation as a function of radius with an average $Q\sim 93$, 
the Quality factor increasing as the size of the galaxies increases as one would expect since with 
larger galaxies the model fitting procedure becomes more reliable.  
The {\sc lensfit} method therefore has an approximate Quality factor of $Q\gs 100$ 
(see Table \ref{Qtable}) which is a factor of at least 
$10$ times better than is required for current weak lensing surveys. 

\section{Conclusion}
\label{Conclusion}
In this paper we have presented the application of the {\sc lensfit} method of Miller et al. (2007) 
to simulated weak lensing data, the Shear TEsting Programme (STEP1 Heymans et al., 2006 and 
STEP2 Massey et al., 2007). The method is a model 
fitting approach to weak lensing shape measurement, the key advancements over other model fitting 
approaches is that it uses realistic galaxy profiles and analytically integrates over the 
position and 
amplitude of the model 
by doing the fitting procedure in Fourier space. Furthermore we use a Bayesian shear 
estimation method which can take into account any bias in a fully self-contained way by using 
a prior ellipticity distribution. In this paper we 
have shown how to estimate the prior distribution from data using an iterative approach which 
we have shown to be stable and convergent. By using this on the STEP1 simulation we have shown that 
this yields a prior distribution which is a good representation 
of the true intrinsic ellipticity distribution. We use the model fitting method to find the full 
posterior probability distribution in ellipticity and then use the Bayesian approach to estimate 
the shear from this distribution.

This method then, should yield a very small bias in the estimated shear. Furthermore it is a fast 
fitting method which takes approximately $1$ second per galaxy (on a 1GHz CPU) 
to find the full posterior probability in ellipticity and is trivially 
parallelisable by assigning one galaxy per CPU. 

The STEP simulations parameterise the ability of a method to measure shear by fitting a linear 
function to the difference between the input (true) shear $\gamma^T$
and the measured shear $\gamma^M$ as a function of the input 
shear $\gamma^M_i-\gamma^T_i= m\gamma^T_i + c$. The values $m$ and $c$ are found for a given method 
which represent any bias in a method and any residual offset in the estimated shear respectively. We
have shown that {\sc lensfit} yields values of 
$m\sim +0.006\pm 0.005$ and $\sigma_c\sim 0.0002$ for the STEP1 
simulations. The variance of $c$ represents the stability of a methods estimation of shear 
to PSF variation. 
This is the smallest combined bias and variance 
for any method, and the smallest bias for any method which has a linear response to the input shear.  

By applying the method to the STEP2 simulations we again found that the bias 
$m\sim 0.002\pm 0.02$ and offset $c=-0.0007$ 
were very small and that the method performed very well in comparison 
to the methods presented in the STEP2 publication. Furthermore when the galaxy sample is split into 
magnitude and size bins, the bias and offset improve over a certain ranges 
since the intrinsic ellipticity prior varies 
as a function of these parameters. By recalculating the prior distribution in each bin the 
intrinsic distribution used 
is a better representation of the galaxies' true ellipticity distribution in that 
bin than if a global average prior is used. The bias was found to be $|m|<0.02$ over magnitudes
$18$ -- $20$ and sizes of galaxy from $0.4$ -- $1.2$ arcseconds. The offset only deviated from $c=0$ 
in the magnitude and size bins where the number of galaxies was $\ls 100$ in which case there 
were too few galaxies to accurately estimate the intrinsic ellipticity prior. However 
this problem will not arise in real surveys since the number of available galaxies will be 
many orders 
of magnitude larger than that in the STEP2 simulation, meaning that any magnitude/size bin will have 
a sufficient number of galaxies to estimate the prior. 

These small biases surpass the predicted requirement for future weak lensing surveys. Amara \& 
Refregier (2007) set a requirement for the DUNE weak lensing concept that any bias 
in shape measurement $m$ needs to be be $\delta m\ls 5\times 10^{-3}$. Kitching et al. (2008) 
present a similar required accuracy of $\delta m\ls 8\times 10^{-3}$ for 
dark energy parameters to remain unbiased. Furthermore, if the 
shape measurement bias is marginalised over as part of the parameter estimation then this 
requirement relaxes to an error on the bias of $\Delta m\ls 10^{-2}$. Thus 
we have shown in this paper that {\sc lensfit} has the potential to negate the concern that 
shape measurement bias may dominate weak lensing systematics.

Going beyond the $m$ and $c$ parameterisation we defined a Quality factor $Q$, 
which quantifies whether the variation in $\gamma^M-\gamma^T$ is purely 
statistical, due to the finite number of galaxies, or whether it is due to some bias in the method. A
$Q=1000$ is where the variance is entirely statistical and $Q\sim 10$ is the limit of current methods 
analysed in the STEP publications. We have shown that using the STEP2 simulation that {\sc lensfit} 
has a Quality factor of $Q\gs 100$, approximately $10$ times better than is required by current
surveys. 

To summarise the main conclusions; 

\begin{itemize} 
\item 
Using the STEP1 simulations we find a bias of $m\sim +6\times 10^{-3}$ 
and a variation in the shear offset 
$\sigma_c\sim 2\times 10^{-4}$. These are some of 
the smallest values for any shape measurement method. 
\item
Using the STEP2 simulations we find a bias of $m\sim 2\times 10^{-3}$ and a shear offset of 
$c\sim -7\times 10^{-4}$, this is the smallest bias of any published method. Furthermore these values do not substantially vary when the shear values from images with highly elliptical PSF's 
are removed suggesting any variation is statistical. 
\item
By analysing the STEP2 simulations as function of size and magnitude the bias and offset 
over a certain range can \emph{improve} relative 
to those found using the entire population as a whole. This is due to the intrinsic 
ellipticity prior's variation as a function of size and magnitude being correctly characterised. 
\item 
We generalise the Quality factor from Bridle et al. (2008) for an arbitrary simulation 
and show that using STEP2 {\sc lensfit} has an average $Q\gs 100$ which is 
at least a factor of $10$ times larger than current methods and the accuracy required 
by current surveys.
\end{itemize}

In a real survey there are a number of sophistications which the STEP simulations do not include. 
None of these should present an insurmountable problem to this method. The PSF will vary as a function
of position, but given a large enough number of stars in each region this can be determined. Currently
we reject any close pairs of galaxies when two or more galaxies lie in the same postage stamp, this 
could be improved so that for pairs in which there is one high signal-to-noise galaxy and one 
very low signal-to-noise galaxy the pair is kept. In cases of multiple exposures the posterior 
probability for each galaxy and each exposure may be combined in an optimal way. 
In other respects the STEP simulations 
are more difficult to analyse using this method than in a real survey, for example 
our assumption that the prior intrinsic ellipticity distribution is  
centred on zero is not true in the STEP simulations since the 
ellipticity is constant across the whole image. 
In reality, where the mean shear across an image should be zero, the 
assumption of a zero-centred prior will be a good representation 
of this distribution.   

The {\sc lensfit} method outperforms the majority of other shape 
measurement methods since it uses realistic 
galaxy profiles and crucially uses a Bayesian method to remove bias. The accuracy with which we have 
shown the method to reach on simulated data sets surpasses the level which current surveys 
require and 
gives confidence that future weak lensing surveys which use such a technique will not be limited 
by the ability to measure the shapes of galaxies.  

\section*{Acknowledgments}
TDK is supported by the Science and Technology Facilities Council,
research grant number E001114. CH is supported by the European Commission Programme 
in the framework of the Marie Curie Fellowship under contract MOIF-CT-2006-21891. 
We thank Richard Massey and the STEP collaboration for making the 
STEP simulations publically available.  
We thank Adam Amara, Sarah Bridle, Konrad Kuijken, Alexandre Refregier and all members of the GREAT08 
team for insightful discussions.


\end{document}